\documentclass[physrev,reprint,superscriptaddress,bibnotes]{revtex4-2}  

\usepackage[T1]{fontenc}
\usepackage[utf8]{inputenc} 
\usepackage[australian]{babel}

\usepackage[a4paper,centering,hmargin=1.75cm,vmargin=2cm]{geometry} 
\usepackage{amsmath,amssymb,graphicx,bm,microtype} 
\usepackage[colorlinks,allcolors=blue!50!black]{hyperref} 
\usepackage[all]{hypcap}
\usepackage{cleveref,braket,siunitx}
\usepackage[version=4]{mhchem}
\usepackage{lipsum}
\usepackage{xcolor}
\usepackage{amssymb}
\usepackage{dsfont}
\usepackage{textgreek}

\usepackage{tikz}

\begin{document}

\title{Universal Quantum Gate Set for Gottesman-Kitaev-Preskill Logical Qubits}

\author{V.~G.~Matsos}
\email{vmat0922@uni.sydney.edu.au}
\affiliation{School of Physics, University of Sydney, NSW 2006, Australia}
\affiliation{ARC Centre of Excellence for Engineered Quantum Systems, University of Sydney, NSW 2006, Australia}

\author{C.~H.~Valahu}
\affiliation{School of Physics, University of Sydney, NSW 2006, Australia}
\affiliation{ARC Centre of Excellence for Engineered Quantum Systems, University of Sydney, NSW 2006, Australia}
\affiliation{Sydney Nano Institute, University of Sydney, NSW 2006, Australia}

\author{M.~J.~Millican}
\affiliation{School of Physics, University of Sydney, NSW 2006, Australia}
\affiliation{ARC Centre of Excellence for Engineered Quantum Systems, University of Sydney, NSW 2006, Australia}

\author{T.~Navickas}
\email{Current address: Q-CTRL, Sydney, NSW Australia}
\affiliation{School of Physics, University of Sydney, NSW 2006, Australia}
\affiliation{ARC Centre of Excellence for Engineered Quantum Systems, University of Sydney, NSW 2006, Australia}

\author{X.~C.~Kolesnikow}
\affiliation{School of Physics, University of Sydney, NSW 2006, Australia}
\affiliation{ARC Centre of Excellence for Engineered Quantum Systems, University of Sydney, NSW 2006, Australia}

\author{M.\,J.~Biercuk}
\affiliation{School of Physics, University of Sydney, NSW 2006, Australia}
\affiliation{ARC Centre of Excellence for Engineered Quantum Systems, University of Sydney, NSW 2006, Australia}

\author{T.~R.~Tan}
\email{tingrei.tan@sydney.edu.au}
\affiliation{School of Physics, University of Sydney, NSW 2006, Australia}
\affiliation{ARC Centre of Excellence for Engineered Quantum Systems, University of Sydney, NSW 2006, Australia}
\affiliation{Sydney Nano Institute, University of Sydney, NSW 2006, Australia}

\begin{abstract}
The realisation of a universal quantum computer at scale promises to deliver a paradigm shift in information processing, providing the capability to solve problems that are intractable with conventional computers. A key limiting factor of realising fault-tolerant quantum information processing (QIP) is the large ratio of physical-to-logical qubits that outstrip device sizes available in the near future. An alternative approach proposed by Gottesman, Kitaev, and Preskill (GKP)~\cite{GKP2001} encodes a single logical qubit into a single harmonic oscillator, alleviating this hardware overhead in exchange for a more complex encoding. Owing to this complexity, current experiments with GKP codes have been limited to single-qubit encodings and operations. Here, we report on the experimental demonstration of a universal gate set for the GKP code, which includes single-qubit gates and -- for the first time -- a two-qubit entangling gate between logical code words. Our scheme deterministically implements energy-preserving quantum gates on finite-energy GKP states encoded in the mechanical motion of a trapped ion. This is achieved by a novel optimal control strategy that dynamically modulates an interaction between the ion's spin and motion. We demonstrate single-qubit gates with a logical process fidelity as high as 0.960 and a two-qubit entangling gate with a logical process fidelity of 0.680. We also directly create a GKP Bell state from the oscillators' ground states in a single step with a logical state fidelity of 0.842. The overall scheme is compatible with existing hardware architectures, highlighting the opportunity to leverage optimal control strategies as a key accelerant towards fault tolerance.
\end{abstract}

\maketitle

\begin{figure*}[t!]
    \centering
    \includegraphics[]{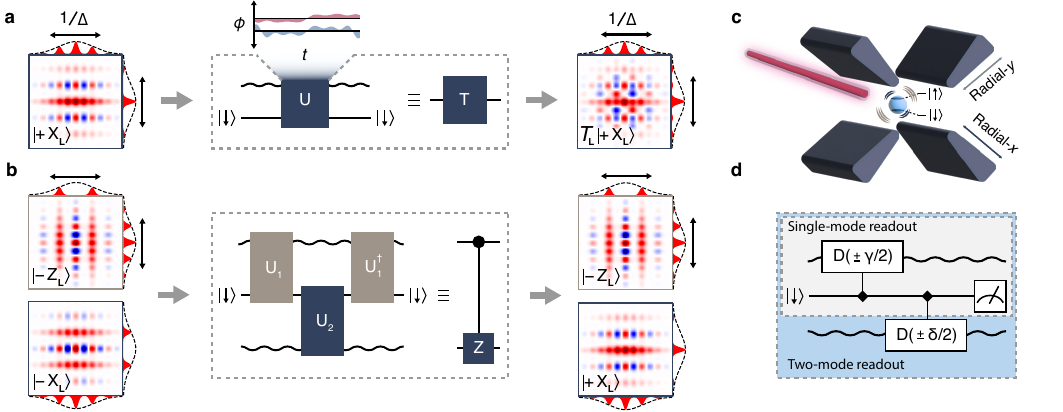}
    \caption{\textbf{Implementation of a universal gate set in an ion trap with logical GKP states.} 
    \textbf{a}, Single-qubit gates are implemented with a unitary, $\hat{U}$. Shown is an example of a $\hat{T}_\mathrm{L}$ gate acting on a $\ket{+X_\mathrm{L}}$ initial GKP state. The Wigner function of the initial and final GKP states are plotted.
    \textbf{b}, A two-qubit controlled-Z gate is implemented with a sequence of three sequential unitaries, $\hat{U}_1$, $\hat{U}_2$ and $\hat{U}_1^\dagger$. 
    The unitaries of the single-qubit and two-qubit gates are obtained from state-dependent force interactions with numerically optimised phases (see the inset of $\hat{U}$ in \textbf{a}). 
    All gates are deterministic and preserve the finite-energy envelope of approximate GKP states, which can be seen from the preservation of the envelope of the final state. The envelope is characterised by the envelope parameter, $\Delta$.
    \textbf{c}, The logical GKP states are encoded in a single trapped ion's radial-$x$ and -$y$ motional modes with frequencies $\{\omega_x, \omega_y\} = 2\pi \times \{1.33, 1.51\}$~MHz. Laser interactions induce spin-dependent forces that couple the ancillary spin and motion.
    \textbf{d}, Single- or two-mode readout of logical GKP Pauli expectation values. The measurement is obtained by readout of the spin in the $\hat{\sigma}_z$ basis after transferring information from the motional modes to the ancilla using displacements conditioned on the ancilla state in the $\hat{\sigma}_x$ basis (black diamond). The values of the displacements, $\gamma$ and $\delta$, allow one to choose the basis of the Pauli measurement.}
    \label{fig:GPExp}
\end{figure*}

Bosonic quantum computing leverages the infinite-dimensional Hilbert spaces of harmonic oscillators to encode error-correctable logical qubits~\cite{Lloyd1999, Braunstein2005}. This encoding strategy can be implemented on existing devices without the need for significant physical redundancy, offering a hardware-efficient avenue for fault-tolerant Quantum Information Processing (QIP).
Among the various bosonic error-correction schemes that have been proposed~\cite{Mirrahimi2014,Michael2016,Grimsmo2020}, the Gottesman, Kitaev, and Preskill (GKP) encoding~\cite{GKP2001} 
provides in-principle superior error-correction performance against dominant hardware decoherence channels~\cite{Albert2018}. This, coupled with the relative ease of implementing Pauli and stabiliser operations, has made it an attractive option for realising bosonic error correction across various physical platforms, including electromagnetic modes in microwave cavities~\cite{Campagne-Ibarcq2020,Eickbusch2022,Sivak2023,Kudra2022,Lachance2023}, optical photons~\cite{konno2024}, and in the mechanical oscillation of trapped ions~\cite{Flhmann2019,Neeve2022,Matsos2024}.
There has also been significant progress in stabilising GKP code words, both using dissipative~\cite{Neeve2022, Lachance2023} and measurement-based~\cite{Campagne-Ibarcq2020, Sivak2023} error-correction. A key outstanding requirement is the realisation of a universal gate set for the GKP code, extending beyond logical single-qubit unitaries~\cite{Flhmann2019}. 

This challenge arises for two primary reasons. The first is the difficulty of engineering high-quality controls to realise an entangling interaction directly between bosonic modes or otherwise through a high-quality entanglement mediator. Another is the need to consider the limitations placed on physically realisable GKP states as they diverge from idealisations~\cite{Royer2020}. Ideal GKP states are unphysical due to their infinite energy, which is evident from their infinite periodic phase space structure. They can be made physical by applying a Gaussian damping envelope that bounds the energy of the state (see Fig.~\ref{fig:GPExp}a-b). Crucially, applying the unitaries designed for logical operations on ideal GKP states to finite-energy GKP code words distorts their Gaussian envelope and causes errors. These errors may be corrected in states that are sufficiently close to the ideal code words at the cost of additional rounds of error correction. There exist several proposals to reduce this overhead~\cite{Eickbusch2022,Royer2020,Hastrup2021ir,Rojkov2023}, but experimental demonstration has been limited to single-qubit operations through a dissipative strategy~\cite{Neeve2022}.

Here, we demonstrate the first realisation of a universal logical gate set for GKP qubits. Our gate set is coherent, deterministic, and designed to avoid code word distortion of finite-energy GKP states. The key enabler of our approach is an optimal control strategy that implements a highly tunable spin-boson interaction via numerically optimised time-domain modulations of experimentally accessible laser fields~\cite{boulder_opal1}. This interaction couples a high-quality ancillary two-level spin and two bosonic modes of a trapped ion (see Fig.~\ref{fig:GPExp} for a visual summary). The spin's excellent coherence properties enable it to function as a nonlinear element for implementing logical GKP operations, and as a mediator of bosonic entanglement with minimal decoherence. We implement a gate set that is made up of the single-qubit gates $\{\hat{R}_\mathrm{L}^X(-\pi/2), \hat{R}_\mathrm{L}^Z(-\pi/2), \hat{T}_\mathrm{L}\}$ and the two-qubit controlled-Z ($\mathrm{CZ}_\mathrm{L}$) gate, which contains all necessary operations for universal quantum computation~\cite{Barenco1995, Bremner2002}. We also demonstrate the preparation of a GKP Bell state from vacuum in a single step, providing an alternative route to universality when paired with our single-qubit gates and measurements~\cite{Raussendorf2001}. We perform logical Quantum Process Tomography (QPT) and logical Quantum State Tomography (QST) to characterise the logical gate set and the Bell state, and we discuss strategies to mitigate identified error sources.

We perform experiments with a single \ce{^{171}Yb^+} ion confined in a room-temperature Paul trap with favourable coherence in both the ancillary qubit and the bosonic modes (see Ref.~\cite{Valahu2023} for details on the experimental setup). The ancillary qubit has a measured $T_2^*$ coherence time of 8.7~s~\cite{Tan2023}; the motional coherence time of an equal superposition of the Fock states, $\ket{0}$ and $\ket{1}$, is measured as $\sim50$~ms~\cite{Matsos2024}; and the heating rate is 0.2 quanta per second. Coherent phase-modulated spin-motion interactions are implemented by a $\SI{355}{nm}$ pulsed laser through stimulated Raman transitions following common routines in the trapped-ion community. 

The control required to prepare and manipulate approximate GKP code words is provided by laser-driven state-dependent forces (SDF) that couple an ancillary spin and bosonic mode. The Hamiltonian for the SDF is described by~\cite{Wineland1998} 
\begin{align}
    \hat{H}_j^\mathrm{SDF}(t) &= \frac{\Omega_j}{2}\hat{\sigma}_{\phi_\mathrm{s}(t)}\left( \hat{a}_j^\dagger e^{-i\phi_\mathrm{m}(t)} + \hat{a}_j e^{i\phi_\mathrm{m}(t)} \right),
    \label{Eq:ControlHamiltonian}
\end{align}
where $\Omega_j=2\pi \times2.4$~kHz is the Rabi rate on bosonic mode $j$, with annihilation operator $\hat{a}_j$; $\phi_\mathrm{s}$ and $\phi_\mathrm{m}$ are independently controllable phases associated with the spin and bosonic mode. The operator $\hat{\sigma}_{\phi_\mathrm{s}(t)} = \hat{\sigma}_x \cos{\phi_\mathrm{s}(t)} + \hat{\sigma}_y \sin{\phi_\mathrm{s}(t)}$ is expressed in terms of Pauli matrices $\hat{\sigma}_x,\hat{\sigma}_y$ that act on the spin state of the ion. The non-commutativity of the modulated SDF interaction provides the requisite non-linearity to prepare and control GKP states~\cite{Matsos2024}. Alternating the SDF interactions between two bosonic modes (see Fig.~\ref{fig:GPExp}b) enables entangling operations between them. We achieve state preparation and logical operations exclusively through modulation of the phases $\phi_\mathrm{s}(t)$ and $\phi_\mathrm{m}(t)$ in $\hat{H}_j^\mathrm{SDF}(t)$. 
The waveforms for these modulations are numerically optimised to prevent distortions of the finite-energy code words. Specifically, the cost function for both single-qubit and two-qubit gates minimises the average error with which a phase-modulated SDF interaction maps a set of finite energy input states to the target corresponding output states~\cite{Eickbusch2022}. The cost function for the preparation of the Bell state minimises the state overlap infidelity with respect to the target finite-energy state. The result of these optimisations is a library of waveforms that can be sequenced in order to enact a desired quantum circuit.

\begin{figure}[t]
    \centering
   \includegraphics[]{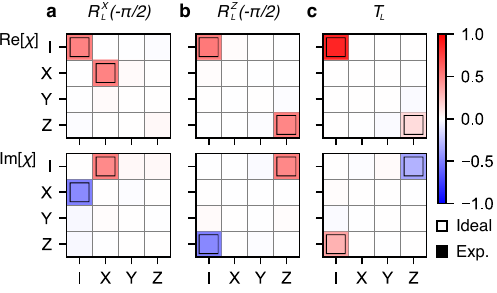}
    \caption{\textbf{Quantum process tomography (QPT) of single-qubit gates.} 
    The real (top) and imaginary parts (bottom) of the $\chi$ matrix obtained from QPT are shown for the \textbf{a},~$\hat{R}^X_\mathrm{L}(-\pi/2)$, \textbf{b},~$\hat{R}^Z_\mathrm{L}(-\pi/2)$ and \textbf{c},~$\hat{T}_\mathrm{L} = \hat{R}^Z_\mathrm{L}(\pi/4)$ gates. The non-zero values of the ideal chi matrix, $\chi_{\mathrm{id}}$, are represented by thick coloured borders around the experimental data. From the experimental $\chi$ matrices, we obtain process fidelities of $\bar{\mathcal{F}}_\mathrm{proc.} = \{0.943, 0.960,0.959\}$. The fidelity is primarily limited by motional dephasing. The decoherence-free gate operations, $\hat{U}$, are limited by imperfections in the pulse design, with each gate having an \textit{average gate fidelity} ${\mathcal{F}}_\mathrm{gate} > 0.998$ (see the SI for more detail).}
    \label{fig:SQ_data}
\end{figure}

In the first experiment, we implement and fully characterise a logical single-qubit (SQ) gate set. The experimental sequence consists of three steps: \textit{state preparation}, \textit{gate application}, and \textit{logical readout}. \textit{State preparation} creates GKP states from vacuum in the radial-$x$ motional mode using a similar optimised state-preparation protocol as detailed in Ref.~\cite{Matsos2024}, augmented to suppress previously neglected errors from higher-order Hamiltonian terms. Numerically optimised waveforms are independently generated to prepare different approximate logical GKP states, including $\ket{+X_\mathrm{L}}$, $\ket{+Y_\mathrm{L}}$, $\ket{+Z_\mathrm{L}}$ and $\ket{-Z_\mathrm{L}}$, with durations between 651 and \SI{821}{\micro\second}. 
The approximate nature of the states is parameterised through their envelope parameter $\Delta$, and the ideal GKP states are retrieved in the limit of infinite squeezing (see Supplemental Information (SI) for details). 
Here and throughout, we use the notation $\ket{\pm P_\mathrm{L}}$ to denote a finite-energy approximate $\pm 1$ eigenstate of the logical GKP Pauli $\hat{P}$ operator. 
Our prepared states have a target envelope parameter between 7.9 and \SI{8.9}{dB}. During \textit{gate application}, $\hat{H}^\mathrm{SDF}_{x}(t)$ is applied with the waveform corresponding to the desired gate, with durations ranging from 196 to \SI{339}{\micro s}. Mid-circuit measurements are performed after state preparation and gate application to remove residual spin-boson entanglement (see supplemental Fig.~\ref{fig:circuits}). Additionally, the experiment only proceeds when the spin is determined to be in the down state $\ket{\downarrow}$ via projective measurement. \textit{Logical readout} is achieved by applying an SDF pulse that transfers logical information from the GKP state to the ancilla, which can then be measured using state-dependent fluorescence (see Fig.~\ref{fig:GPExp}d). This measurement probes the real part of the expectation value of a displacement operator, $\operatorname{Re}\langle \hat{D}(\gamma)\rangle $, where $\hat{D}(\gamma) = e^{\gamma \hat{a}^\dagger - \gamma^* \hat{a}}$ denotes the bosonic displacement operator. 

The experimental sequence outlined above describes one step of a QPT routine on the logical single-qubit gates. By preparing a range of input states and decoding the input and corresponding output states, QPT characterises a completely positive trace-preserving map, $\mathcal{E}$, that relates the reconstructed logical input and output density matrices via $\hat{\rho}_\mathrm{L}^\mathrm{out} = \mathcal{E}(\hat{\rho}_\mathrm{L}^\mathrm{in}) = \sum_{m,n} \chi_{m,n} \hat{E}_m \hat{\rho}_\mathrm{L}^{\text{in}} \hat{E}_n^\dagger$, where $\hat{E}_m$ are Pauli operators. The complex matrix $\chi$ fully describes $\mathcal{E}$, and can be used to determine the process fidelity, $\bar{\mathcal{F}}_\mathrm{proc.} = \textrm{Tr}(\chi\chi_\mathrm{id})$, where $\chi_\mathrm{id}$ describes the ideal action of the gate on two-level systems.

We use the stabiliser sub-system decomposition (SSSD) formalism~\cite{Shaw2024} to efficiently retrieve the logical information contained in the finite-energy GKP qubits. This approach addresses the shortcomings of measuring finite-energy GKP states via the conventional usage of the logical Pauli operators, $\{\langle\hat{X}_\mathrm{L}\rangle, \langle\hat{Y}_\mathrm{L}\rangle, \langle\hat{Z}_\mathrm{L}\rangle \}$. The SSSD may be used to provide a more accurate fidelity estimate by measuring the expectation value of \textit{logical Pauli measurement operators} (henceforth referred to as Pauli measurement operators for brevity), $\{\langle \hat{X}_\mathrm{m}\rangle,\langle \hat{Y}_\mathrm{m}\rangle,\langle \hat{Z}_\mathrm{m}\rangle\}$. Each expectation value of the Pauli measurement operators is obtained by a weighted sum of displacements on the GKP lattice (see SI for details)~\cite{Shaw2024a}. In practice, the truncation of this sum can be chosen to obtain adequate accuracy of the fidelity. The reconstructed $\chi$ matrices obtained from QPT through SSSD for the single-qubit gate set $ \{ \hat{R}^X_\mathrm{L}(-\pi/2), \hat{R}^Z_\mathrm{L}(-\pi/2), \hat{T}_\mathrm{L}= \hat{R}^Z_\mathrm{L}(\pi/4)\} $ are shown in Fig.~\ref{fig:SQ_data}. From this, we compute logical process fidelities of $\bar{\mathcal{F}}_\mathrm{proc.} = \{0.943, 0.960,0.959 \}$. 

\begin{figure}[t]
    \centering
   \includegraphics[]{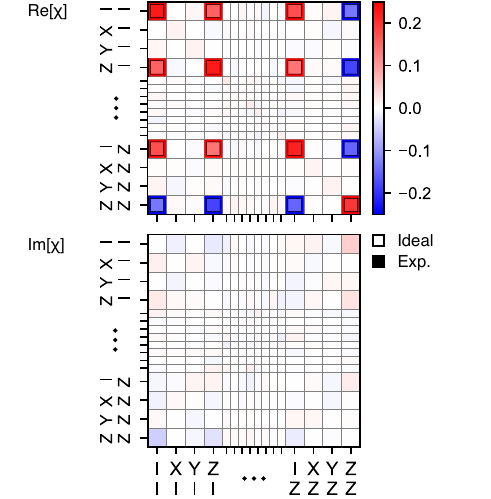}
    \caption{\textbf{Quantum process tomography (QPT) of a two-qubit logical entangling gate.} 
    The real (top) and imaginary (bottom) parts of the $\chi$ matrix obtained from QPT are shown for the controlled-Z gate. The non-zero values of the ideal chi matrix, $\chi_{\mathrm{id}}$, are represented by thick coloured borders around the experimental data. Elements of the $\chi$ matrix that are near-zero are plotted with smaller rows and columns, and the omitted labels ($\boldsymbol{\cdot \cdot \cdot}$) are $\{IX, XX, YX, ZX, IY, XY, YY, ZY\}$. The imaginary part of the experimental $\chi$ matrix is small, in agreement with the vanishing ideal matrix elements of $\chi_{\mathrm{id}}$. The process fidelity obtained from the experimental $\chi$ matrix is $\bar{\mathcal{F}}_\mathrm{proc.} = 0.680$, with errors dominated by motional dephasing. The decoherence-free gate operation, $\hat{U}_\mathrm{CZ}$, is limited by pulse-design imperfections and has an \textit{average gate fidelity} ${\mathcal{F}}_\mathrm{gate} = 0.990$ (see the SI for more detail).}
    \label{fig:TQ_data}
\end{figure}

In the second experiment, we implement and characterise a two-qubit $\mathrm{CZ}_\mathrm{L}$ entangling gate by interleaving SDF interactions between bosonic modes. In particular, the $\mathrm{CZ}_\mathrm{L}$ gate is decomposed into three sequential unitaries, $\hat{U}_\mathrm{CZ} = \hat{U}_1^\dagger \hat{U}_2 \hat{U}_1$ (see Fig.~\ref{fig:GPExp}b), each involving a single bosonic mode, leading to a total gate duration of \SI{993}{ \micro s};  $\hat{U}_1$ and $\hat{U}_2$ are generated by applying $\hat{H}^\mathrm{SDF}_y$ and $\hat{H}^\mathrm{SDF}_x$, respectively. Decomposing the $\mathrm{CZ}_\mathrm{L}$ operation into three individual spin-boson interactions offers two significant advantages: first, this reduces the complexity of the experimental implementation by requiring fewer calibration steps; second, it increases the computational efficiency of control-waveform design, since expensive tasks in the numerical optimisation need only consider a reduced Hilbert space. Computational efficiency is further increased by constraining the first and third unitary operations to be inverse of one another. 

We characterise the two-qubit $\mathrm{CZ}_\mathrm{L}$ gate using QPT to reconstruct the $\chi$ matrix and determine the process fidelity. The experimental sequence is similar to the single-qubit experiment described above. First, \textit{state preparation} uses numerically optimised preparation waveforms to sequentially initialise GKP states in the radial-$y$ and radial-$x$ modes. \textit{Gate application} implements the $\mathrm{CZ}_\mathrm{L}$ operation by applying three sequential optimised SDF pulses, as described above. Similar to the single-qubit case, \textit{logical readout} uses two sequential SDF pulses to transfer information from both bosonic modes to the ancilla (see Fig.~\ref{fig:GPExp}d). This allows simultaneous measurement of both modes of the form $\textrm{Re}\langle {\hat{D}}(\gamma) \otimes \hat{D}(\delta)\rangle$. Setting $\gamma$ and $\delta$ to points on the GKP square lattice results in joint expectation values of the logical Pauli operators (e.g. $\langle \hat{X}_{L} \otimes \hat{Z}_{L} \rangle$), which inform the $\chi$ matrix calculation. Here, we only measure the expectation value of the logical Pauli operators instead of the Pauli measurement operators to reduce the total number of required measurements. This introduces a non-negligible error that we discuss below. The reconstructed $\chi$ matrix for the two-qubit logical $\mathrm{CZ}_\mathrm{L}$ gate is shown in Fig.~\ref{fig:TQ_data}, from which we obtain a logical process fidelity of $\bar{\mathcal{F}}_\mathrm{proc.} = 0.680$. The experimental and ideal $\chi$ matrices show good qualitative agreement: the dominant elements of the matrix are real and are located in the Pauli basis labels $\{II, ZI, IZ, ZZ\}$, and the imaginary part of the experimental $\chi$ matrix remains near zero. The infidelity is due to known experimental errors, described below. 

\begin{figure}[t]
    \centering
   \includegraphics[]{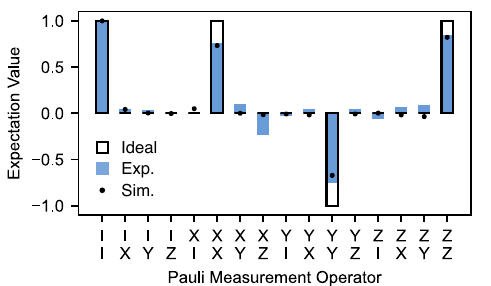}
    \caption{\textbf{Quantum state tomography (QST) of a maximally entangled GKP logical Bell state}. The Bell state $\ket{\Phi^+_\mathrm{L}} = \left(\ket{+Z_\mathrm{L}, +Z_\mathrm{L}} +\ket{-Z_\mathrm{L}, -Z_\mathrm{L}}\right)/\sqrt{2}$ is prepared from vacuum using an optimised control pulse. Experimental data (shaded bars) are expectation values of Pauli measurement operators obtained from QST. Black bars represent corresponding expectation values of an ideal Bell state in two-level systems. We determine a logical state fidelity of $\bar{\mathcal{F}} =0.842$. Black circles are obtained from numerical simulations of the experimental pulse sequence with motional dephasing, corresponding to a logical state fidelity of $\bar{\mathcal{F}} =0.807$. The experiment does not measure $\langle \hat{I}_\mathrm{m}\hat{I}_\mathrm{m}\rangle$, which is set to 1. The decoherence-free operation, $\hat{U}_\mathrm{Bell}$, is limited by imperfections in the pulse design and has a state fidelity of 0.984 (see SI for details).}
    \label{fig:bell_data}
\end{figure}

In the third experiment, we again use an optimal control strategy to directly prepare a GKP Bell state from vacuum in a single step, representing an efficient method for GKP resource-state synthesis~\cite{Bourassa2021}. This direct preparation removes the usual overhead of state preparation followed by an entangling gate to generate a Bell state.
We prepare the logical Bell state, $\ket{\Phi^+_\mathrm{L}} = \left(\ket{+Z_\mathrm{L}, +Z_\mathrm{L}} +\ket{-Z_\mathrm{L}, -Z_\mathrm{L}}\right)/\sqrt{2}$, from the initial ground state, $\ket{\downarrow, 0,0}$. Similarly to $\mathrm{CZ}_\mathrm{L}$, the unitary operation $\hat{U}_\mathrm{Bell}$ is decomposed into three numerically optimised unitaries applied in sequence, $\hat{U}_\mathrm{Bell} = \hat{U}_3 \hat{U}_2 \hat{U}_1$, with a total duration of \SI{1.86}{ \milli s}.

The Bell state is characterised using QST by measuring the two-mode Pauli measurement operators, e.g., $\langle \hat{X}_\mathrm{m} \otimes \hat{Z}_\mathrm{m}\rangle$.
Experimental expectation values of these operators are shown in Fig.~\ref{fig:bell_data}. These measurements are used to construct a logical density matrix, $\hat{\rho}_{\mathrm{L},\mathrm{exp}}$, which contains the logical information encoded in the state (see SI). Performing \textit{logical} state tomography through the SSSD in this way requires a vastly reduced set of measurements when compared to the number of measurements required to perform full-state tomography for a two-mode bosonic state. From this strategy, we determine a logical state fidelity~\cite{Kolesnikow2024}, $\bar{\mathcal{F}} = \bra{\Phi^+} \hat{\rho}_{\mathrm{L},\mathrm{ exp}} \ket{\Phi^+}$, of $0.842$.

Investigation of the impact of various noise mechanisms in our experiment reveals that the dominant error source is motional dephasing originating from fluctuations in the trapping potential. The two-qubit-gate experiment also suffers from the non-negligible logical readout errors that result from measurements of the logical Pauli operators on finite-energy states. This can be mitigated by measuring the Pauli measurement operators at the cost of more experimental measurements. In the Bell state experiment, the state fidelity suffers additional infidelities from thermal noise due to imperfect bosonic ground-state initialisation. A comprehensive analysis of the dominant errors is given in the SI.  

In summary, we have demonstrated a universal gate set and direct Bell state generation for finite-energy GKP qubits encoded in the mechanical motion of a trapped ion. Our approach uses numerically optimised dynamically modulated light-atom interactions to implement coherent operations that are made to preserve the finite-energy envelope of GKP code words. These control pulses take advantage of the excellent coherence properties of the spin present in the trapped-ion system, which serves as a resource-efficient nonlinear element to enable high controllability and as a high-quality mediator of bosonic entanglement. We characterised these operations with an efficient multi-mode logical tomography procedure tailored for finite-energy GKP qubits.

Our investigations have identified straightforward hardware improvements to suppress infidelities. Errors arising from motional dephasing can be mitigated by reducing the dissipative rate in the hardware while increasing the speed of state preparation and logical gates. A faster speed can be achieved by increasing our light-atom coupling strength, $\Omega_j$, which is modest compared to state-of-the-art trapped ion platforms~\cite{Schafer2018}. We estimate through numerical simulation that a 10$\times$ improvement of the coupling strength and increasing the motional coherence time to $\SI{200}{ms}$ suppresses the dephasing error to below $5\times10^{-3}$. Similarly, thermal noise in the bosonic modes can also be suppressed by improving the bosonic-mode initialisation protocol~\cite{Chou2017}. 

Looking forward, an exciting prospect is to incorporate the tools developed here with existing trapped-ion hardware, such as linear Coulomb crystals~\cite{Postler2022,chen2023arXiv}, two-dimensional ion arrays~\cite{Warring2020,Kiesenhofer2023,Guo2024,Lysne2024}, quantum charge-coupled  devices~\cite{Wineland1998,Moses2023}, and micro-Penning traps~\cite{Jain2024}; these tools are also applicable to controlling motional modes of neutral atoms trapped in optical tweezers~\cite{shaw2023}. In addition, we can also combine our universal gate set with quantum error correction using quantum non-demolition measurements of protected motional modes~\cite{Hou2024}. The demonstrated multi-mode control presents an opportunity to explore exotic multi-mode grid state codes~\cite{Royer2022} or to investigate operations on codes that may exhibit superior error correction performance against dephasing noise, such as rotation-symmetric codes~\cite{Grimsmo2020}. Finally, the excellent spin-boson control developed here may aid the development of new hybrid discrete- and continuous-variable QIP schemes~\cite{Lui2024}, which have the potential to significantly advance quantum information science.

\begin{acknowledgments}

We thank B. Baragiola, N. Mennicucci, M. Stafford, Z. Huang, T. Smith, and R. Harper for fruitful discussions. 

We were supported by the Australian Research Council (FT220100359), the Sydney Horizon Fellowship (TRT), the U.S. Office of Naval Research Global (N62909-24-1-2083), the U.S. Army Research Office Laboratory for Physical Sciences (W911NF-21-1-0003), the U.S. Air Force Office of Scientific Research (FA2386-23-1-4062), Lockheed Martin, the Sydney Quantum Academy (MJM), the University of Sydney Postgraduate Award scholarship (VGM), the Australian Government Research Training Program scholarship (XCK) and H.\ and A.\ Harley.
\end{acknowledgments}

\section*{Data availability}

The experimental data are available upon reasonable request.

\section*{Author contributions}

V.G.M., C.H.V. and T.R.T. conceived the idea. V.G.M developed the SDF numerical optimisation, performed the experiment, and analysed the data. V.G.M., C.H.V, M.J.M, T.R.T., and T.N. contributed to the experimental apparatus, and X.C.K. provided theoretical support. T.R.T. supervised the project. V.G.M., C.H.V., and T.R.T. wrote the manuscript. All authors provided suggestions for the experiment, discussed the results, and contributed to the manuscript.

\bibliography{ref}

\begin{thebibliography}{55}%
\makeatletter
\providecommand \@ifxundefined [1]{%
 \@ifx{#1\undefined}
}%
\providecommand \@ifnum [1]{%
 \ifnum #1\expandafter \@firstoftwo
 \else \expandafter \@secondoftwo
 \fi
}%
\providecommand \@ifx [1]{%
 \ifx #1\expandafter \@firstoftwo
 \else \expandafter \@secondoftwo
 \fi
}%
\providecommand \natexlab [1]{#1}%
\providecommand \enquote  [1]{``#1''}%
\providecommand \bibnamefont  [1]{#1}%
\providecommand \bibfnamefont [1]{#1}%
\providecommand \citenamefont [1]{#1}%
\providecommand \href@noop [0]{\@secondoftwo}%
\providecommand \href [0]{\begingroup \@sanitize@url \@href}%
\providecommand \@href[1]{\@@startlink{#1}\@@href}%
\providecommand \@@href[1]{\endgroup#1\@@endlink}%
\providecommand \@sanitize@url [0]{\catcode `\\12\catcode `\$12\catcode `\&12\catcode `\#12\catcode `\^12\catcode `\_12\catcode `\%12\relax}%
\providecommand \@@startlink[1]{}%
\providecommand \@@endlink[0]{}%
\providecommand \url  [0]{\begingroup\@sanitize@url \@url }%
\providecommand \@url [1]{\endgroup\@href {#1}{\urlprefix }}%
\providecommand \urlprefix  [0]{URL }%
\providecommand \Eprint [0]{\href }%
\providecommand \doibase [0]{https://doi.org/}%
\providecommand \selectlanguage [0]{\@gobble}%
\providecommand \bibinfo  [0]{\@secondoftwo}%
\providecommand \bibfield  [0]{\@secondoftwo}%
\providecommand \translation [1]{[#1]}%
\providecommand \BibitemOpen [0]{}%
\providecommand \bibitemStop [0]{}%
\providecommand \bibitemNoStop [0]{.\EOS\space}%
\providecommand \EOS [0]{\spacefactor3000\relax}%
\providecommand \BibitemShut  [1]{\csname bibitem#1\endcsname}%
\let\auto@bib@innerbib\@empty
\bibitem [{\citenamefont {Gottesman}\ \emph {et~al.}(2001)\citenamefont {Gottesman}, \citenamefont {Kitaev},\ and\ \citenamefont {Preskill}}]{GKP2001}%
  \BibitemOpen
  \bibfield  {author} {\bibinfo {author} {\bibfnamefont {D.}~\bibnamefont {Gottesman}}, \bibinfo {author} {\bibfnamefont {A.}~\bibnamefont {Kitaev}},\ and\ \bibinfo {author} {\bibfnamefont {J.}~\bibnamefont {Preskill}},\ }\bibfield  {title} {\bibinfo {title} {Encoding a qubit in an oscillator},\ }\href {https://doi.org/10.1103/PhysRevA.64.012310} {\bibfield  {journal} {\bibinfo  {journal} {Phys. Rev. A}\ }\textbf {\bibinfo {volume} {64}},\ \bibinfo {pages} {012310} (\bibinfo {year} {2001})}\BibitemShut {NoStop}%
\bibitem [{\citenamefont {Lloyd}\ and\ \citenamefont {Braunstein}(1999)}]{Lloyd1999}%
  \BibitemOpen
  \bibfield  {author} {\bibinfo {author} {\bibfnamefont {S.}~\bibnamefont {Lloyd}}\ and\ \bibinfo {author} {\bibfnamefont {S.~L.}\ \bibnamefont {Braunstein}},\ }\bibfield  {title} {\bibinfo {title} {Quantum computation over continuous variables},\ }\href {https://doi.org/10.1103/PhysRevLett.82.1784} {\bibfield  {journal} {\bibinfo  {journal} {Phys. Rev. Lett.}\ }\textbf {\bibinfo {volume} {82}},\ \bibinfo {pages} {1784} (\bibinfo {year} {1999})}\BibitemShut {NoStop}%
\bibitem [{\citenamefont {Braunstein}\ and\ \citenamefont {van Loock}(2005)}]{Braunstein2005}%
  \BibitemOpen
  \bibfield  {author} {\bibinfo {author} {\bibfnamefont {S.~L.}\ \bibnamefont {Braunstein}}\ and\ \bibinfo {author} {\bibfnamefont {P.}~\bibnamefont {van Loock}},\ }\bibfield  {title} {\bibinfo {title} {Quantum information with continuous variables},\ }\href {https://doi.org/10.1103/RevModPhys.77.513} {\bibfield  {journal} {\bibinfo  {journal} {Rev. Mod. Phys.}\ }\textbf {\bibinfo {volume} {77}},\ \bibinfo {pages} {513} (\bibinfo {year} {2005})}\BibitemShut {NoStop}%
\bibitem [{\citenamefont {Mirrahimi}\ \emph {et~al.}(2014)\citenamefont {Mirrahimi}, \citenamefont {Leghtas}, \citenamefont {Albert}, \citenamefont {Touzard}, \citenamefont {Schoelkopf}, \citenamefont {Jiang},\ and\ \citenamefont {Devoret}}]{Mirrahimi2014}%
  \BibitemOpen
  \bibfield  {author} {\bibinfo {author} {\bibfnamefont {M.}~\bibnamefont {Mirrahimi}}, \bibinfo {author} {\bibfnamefont {Z.}~\bibnamefont {Leghtas}}, \bibinfo {author} {\bibfnamefont {V.~V.}\ \bibnamefont {Albert}}, \bibinfo {author} {\bibfnamefont {S.}~\bibnamefont {Touzard}}, \bibinfo {author} {\bibfnamefont {R.~J.}\ \bibnamefont {Schoelkopf}}, \bibinfo {author} {\bibfnamefont {L.}~\bibnamefont {Jiang}},\ and\ \bibinfo {author} {\bibfnamefont {M.~H.}\ \bibnamefont {Devoret}},\ }\bibfield  {title} {\bibinfo {title} {Dynamically protected cat-qubits: a new paradigm for universal quantum computation},\ }\href {https://doi.org/10.1088/1367-2630/16/4/045014} {\bibfield  {journal} {\bibinfo  {journal} {New J. Phys.}\ }\textbf {\bibinfo {volume} {16}},\ \bibinfo {pages} {045014} (\bibinfo {year} {2014})}\BibitemShut {NoStop}%
\bibitem [{\citenamefont {Michael}\ \emph {et~al.}(2016)\citenamefont {Michael}, \citenamefont {Silveri}, \citenamefont {Brierley}, \citenamefont {Albert}, \citenamefont {Salmilehto}, \citenamefont {Jiang},\ and\ \citenamefont {Girvin}}]{Michael2016}%
  \BibitemOpen
  \bibfield  {author} {\bibinfo {author} {\bibfnamefont {M.~H.}\ \bibnamefont {Michael}}, \bibinfo {author} {\bibfnamefont {M.}~\bibnamefont {Silveri}}, \bibinfo {author} {\bibfnamefont {R.}~\bibnamefont {Brierley}}, \bibinfo {author} {\bibfnamefont {V.~V.}\ \bibnamefont {Albert}}, \bibinfo {author} {\bibfnamefont {J.}~\bibnamefont {Salmilehto}}, \bibinfo {author} {\bibfnamefont {L.}~\bibnamefont {Jiang}},\ and\ \bibinfo {author} {\bibfnamefont {S.}~\bibnamefont {Girvin}},\ }\bibfield  {title} {\bibinfo {title} {New class of quantum error-correcting codes for a bosonic mode},\ }\href {https://doi.org/10.1103/physrevx.6.031006} {\bibfield  {journal} {\bibinfo  {journal} {Physical Review X}\ }\textbf {\bibinfo {volume} {6}},\ \bibinfo {pages} {3} (\bibinfo {year} {2016})}\BibitemShut {NoStop}%
\bibitem [{\citenamefont {Grimsmo}\ \emph {et~al.}(2020)\citenamefont {Grimsmo}, \citenamefont {Combes},\ and\ \citenamefont {Baragiola}}]{Grimsmo2020}%
  \BibitemOpen
  \bibfield  {author} {\bibinfo {author} {\bibfnamefont {A.~L.}\ \bibnamefont {Grimsmo}}, \bibinfo {author} {\bibfnamefont {J.}~\bibnamefont {Combes}},\ and\ \bibinfo {author} {\bibfnamefont {B.~Q.}\ \bibnamefont {Baragiola}},\ }\bibfield  {title} {\bibinfo {title} {Quantum computing with rotation-symmetric bosonic codes},\ }\href {https://doi.org/10.1103/PhysRevX.10.011058} {\bibfield  {journal} {\bibinfo  {journal} {Phys. Rev. X}\ }\textbf {\bibinfo {volume} {10}},\ \bibinfo {pages} {011058} (\bibinfo {year} {2020})}\BibitemShut {NoStop}%
\bibitem [{\citenamefont {Albert}\ \emph {et~al.}(2018)\citenamefont {Albert}, \citenamefont {Noh}, \citenamefont {Duivenvoorden}, \citenamefont {Young}, \citenamefont {Brierley}, \citenamefont {Reinhold}, \citenamefont {Vuillot}, \citenamefont {Li}, \citenamefont {Shen}, \citenamefont {Girvin}, \citenamefont {Terhal},\ and\ \citenamefont {Jiang}}]{Albert2018}%
  \BibitemOpen
  \bibfield  {author} {\bibinfo {author} {\bibfnamefont {V.~V.}\ \bibnamefont {Albert}}, \bibinfo {author} {\bibfnamefont {K.}~\bibnamefont {Noh}}, \bibinfo {author} {\bibfnamefont {K.}~\bibnamefont {Duivenvoorden}}, \bibinfo {author} {\bibfnamefont {D.~J.}\ \bibnamefont {Young}}, \bibinfo {author} {\bibfnamefont {R.~T.}\ \bibnamefont {Brierley}}, \bibinfo {author} {\bibfnamefont {P.}~\bibnamefont {Reinhold}}, \bibinfo {author} {\bibfnamefont {C.}~\bibnamefont {Vuillot}}, \bibinfo {author} {\bibfnamefont {L.}~\bibnamefont {Li}}, \bibinfo {author} {\bibfnamefont {C.}~\bibnamefont {Shen}}, \bibinfo {author} {\bibfnamefont {S.~M.}\ \bibnamefont {Girvin}}, \bibinfo {author} {\bibfnamefont {B.~M.}\ \bibnamefont {Terhal}},\ and\ \bibinfo {author} {\bibfnamefont {L.}~\bibnamefont {Jiang}},\ }\bibfield  {title} {\bibinfo {title} {Performance and structure of single-mode bosonic codes},\ }\href {https://doi.org/10.1103/PhysRevA.97.032346} {\bibfield  {journal} {\bibinfo  {journal} {Phys. Rev. A}\ }\textbf {\bibinfo
  {volume} {97}},\ \bibinfo {pages} {032346} (\bibinfo {year} {2018})}\BibitemShut {NoStop}%
\bibitem [{\citenamefont {Campagne-Ibarcq}\ \emph {et~al.}(2020)\citenamefont {Campagne-Ibarcq}, \citenamefont {Eickbusch}, \citenamefont {Touzard}, \citenamefont {Zalys-Geller}, \citenamefont {Frattini}, \citenamefont {Sivak}, \citenamefont {Reinhold}, \citenamefont {Puri}, \citenamefont {Shankar}, \citenamefont {Schoelkopf}, \citenamefont {Frunzio}, \citenamefont {Mirrahimi},\ and\ \citenamefont {Devoret}}]{Campagne-Ibarcq2020}%
  \BibitemOpen
  \bibfield  {author} {\bibinfo {author} {\bibfnamefont {P.}~\bibnamefont {Campagne-Ibarcq}}, \bibinfo {author} {\bibfnamefont {A.}~\bibnamefont {Eickbusch}}, \bibinfo {author} {\bibfnamefont {S.}~\bibnamefont {Touzard}}, \bibinfo {author} {\bibfnamefont {E.}~\bibnamefont {Zalys-Geller}}, \bibinfo {author} {\bibfnamefont {N.~E.}\ \bibnamefont {Frattini}}, \bibinfo {author} {\bibfnamefont {V.~V.}\ \bibnamefont {Sivak}}, \bibinfo {author} {\bibfnamefont {P.}~\bibnamefont {Reinhold}}, \bibinfo {author} {\bibfnamefont {S.}~\bibnamefont {Puri}}, \bibinfo {author} {\bibfnamefont {S.}~\bibnamefont {Shankar}}, \bibinfo {author} {\bibfnamefont {R.~J.}\ \bibnamefont {Schoelkopf}}, \bibinfo {author} {\bibfnamefont {L.}~\bibnamefont {Frunzio}}, \bibinfo {author} {\bibfnamefont {M.}~\bibnamefont {Mirrahimi}},\ and\ \bibinfo {author} {\bibfnamefont {M.~H.}\ \bibnamefont {Devoret}},\ }\bibfield  {title} {\bibinfo {title} {{Quantum error correction of a qubit encoded in grid states of an oscillator}},\ }\href
  {https://doi.org/10.1038/s41586-020-2603-3} {\bibfield  {journal} {\bibinfo  {journal} {Nature}\ }\textbf {\bibinfo {volume} {584}},\ \bibinfo {pages} {368} (\bibinfo {year} {2020})}\BibitemShut {NoStop}%
\bibitem [{\citenamefont {Eickbusch}\ \emph {et~al.}(2022)\citenamefont {Eickbusch}, \citenamefont {Sivak}, \citenamefont {Ding}, \citenamefont {Elder}, \citenamefont {Jha}, \citenamefont {Venkatraman}, \citenamefont {Royer}, \citenamefont {Girvin}, \citenamefont {Schoelkopf},\ and\ \citenamefont {Devoret}}]{Eickbusch2022}%
  \BibitemOpen
  \bibfield  {author} {\bibinfo {author} {\bibfnamefont {A.}~\bibnamefont {Eickbusch}}, \bibinfo {author} {\bibfnamefont {V.}~\bibnamefont {Sivak}}, \bibinfo {author} {\bibfnamefont {A.~Z.}\ \bibnamefont {Ding}}, \bibinfo {author} {\bibfnamefont {S.~S.}\ \bibnamefont {Elder}}, \bibinfo {author} {\bibfnamefont {S.~R.}\ \bibnamefont {Jha}}, \bibinfo {author} {\bibfnamefont {J.}~\bibnamefont {Venkatraman}}, \bibinfo {author} {\bibfnamefont {B.}~\bibnamefont {Royer}}, \bibinfo {author} {\bibfnamefont {S.~M.}\ \bibnamefont {Girvin}}, \bibinfo {author} {\bibfnamefont {R.~J.}\ \bibnamefont {Schoelkopf}},\ and\ \bibinfo {author} {\bibfnamefont {M.~H.}\ \bibnamefont {Devoret}},\ }\bibfield  {title} {\bibinfo {title} {Fast universal control of an oscillator with weak dispersive coupling to a qubit},\ }\href {https://doi.org/10.1038/s41567-022-01776-9} {\bibfield  {journal} {\bibinfo  {journal} {Nat. Phys.}\ }\textbf {\bibinfo {volume} {18}},\ \bibinfo {pages} {1464} (\bibinfo {year} {2022})}\BibitemShut {NoStop}%
\bibitem [{\citenamefont {Sivak}\ \emph {et~al.}(2023)\citenamefont {Sivak}, \citenamefont {Eickbusch}, \citenamefont {Royer}, \citenamefont {Singh}, \citenamefont {Tsioutsios}, \citenamefont {Ganjam}, \citenamefont {Miano}, \citenamefont {Brock}, \citenamefont {Ding}, \citenamefont {Frunzio}, \citenamefont {Girvin}, \citenamefont {Schoelkopf},\ and\ \citenamefont {Devoret}}]{Sivak2023}%
  \BibitemOpen
  \bibfield  {author} {\bibinfo {author} {\bibfnamefont {V.~V.}\ \bibnamefont {Sivak}}, \bibinfo {author} {\bibfnamefont {A.}~\bibnamefont {Eickbusch}}, \bibinfo {author} {\bibfnamefont {B.}~\bibnamefont {Royer}}, \bibinfo {author} {\bibfnamefont {S.}~\bibnamefont {Singh}}, \bibinfo {author} {\bibfnamefont {I.}~\bibnamefont {Tsioutsios}}, \bibinfo {author} {\bibfnamefont {S.}~\bibnamefont {Ganjam}}, \bibinfo {author} {\bibfnamefont {A.}~\bibnamefont {Miano}}, \bibinfo {author} {\bibfnamefont {B.~L.}\ \bibnamefont {Brock}}, \bibinfo {author} {\bibfnamefont {A.~Z.}\ \bibnamefont {Ding}}, \bibinfo {author} {\bibfnamefont {L.}~\bibnamefont {Frunzio}}, \bibinfo {author} {\bibfnamefont {S.~M.}\ \bibnamefont {Girvin}}, \bibinfo {author} {\bibfnamefont {R.~J.}\ \bibnamefont {Schoelkopf}},\ and\ \bibinfo {author} {\bibfnamefont {M.~H.}\ \bibnamefont {Devoret}},\ }\bibfield  {title} {\bibinfo {title} {Real-time quantum error correction beyond break-even},\ }\href {https://doi.org/10.1038/s41586-023-05782-6} {\bibfield
  {journal} {\bibinfo  {journal} {Nature}\ }\textbf {\bibinfo {volume} {616}},\ \bibinfo {pages} {50} (\bibinfo {year} {2023})}\BibitemShut {NoStop}%
\bibitem [{\citenamefont {Kudra}\ \emph {et~al.}(2022)\citenamefont {Kudra}, \citenamefont {Kervinen}, \citenamefont {Strandberg}, \citenamefont {Ahmed}, \citenamefont {Scigliuzzo}, \citenamefont {Osman}, \citenamefont {Lozano}, \citenamefont {Thol\'en}, \citenamefont {Borgani}, \citenamefont {Haviland}, \citenamefont {Ferrini}, \citenamefont {Bylander}, \citenamefont {Kockum}, \citenamefont {Quijandr\'{\i}a}, \citenamefont {Delsing},\ and\ \citenamefont {Gasparinetti}}]{Kudra2022}%
  \BibitemOpen
  \bibfield  {author} {\bibinfo {author} {\bibfnamefont {M.}~\bibnamefont {Kudra}}, \bibinfo {author} {\bibfnamefont {M.}~\bibnamefont {Kervinen}}, \bibinfo {author} {\bibfnamefont {I.}~\bibnamefont {Strandberg}}, \bibinfo {author} {\bibfnamefont {S.}~\bibnamefont {Ahmed}}, \bibinfo {author} {\bibfnamefont {M.}~\bibnamefont {Scigliuzzo}}, \bibinfo {author} {\bibfnamefont {A.}~\bibnamefont {Osman}}, \bibinfo {author} {\bibfnamefont {D.~P.}\ \bibnamefont {Lozano}}, \bibinfo {author} {\bibfnamefont {M.~O.}\ \bibnamefont {Thol\'en}}, \bibinfo {author} {\bibfnamefont {R.}~\bibnamefont {Borgani}}, \bibinfo {author} {\bibfnamefont {D.~B.}\ \bibnamefont {Haviland}}, \bibinfo {author} {\bibfnamefont {G.}~\bibnamefont {Ferrini}}, \bibinfo {author} {\bibfnamefont {J.}~\bibnamefont {Bylander}}, \bibinfo {author} {\bibfnamefont {A.~F.}\ \bibnamefont {Kockum}}, \bibinfo {author} {\bibfnamefont {F.}~\bibnamefont {Quijandr\'{\i}a}}, \bibinfo {author} {\bibfnamefont {P.}~\bibnamefont {Delsing}},\ and\ \bibinfo {author}
  {\bibfnamefont {S.}~\bibnamefont {Gasparinetti}},\ }\bibfield  {title} {\bibinfo {title} {Robust preparation of wigner-negative states with optimized snap-displacement sequences},\ }\href {https://doi.org/10.1103/PRXQuantum.3.030301} {\bibfield  {journal} {\bibinfo  {journal} {PRX Quantum}\ }\textbf {\bibinfo {volume} {3}},\ \bibinfo {pages} {030301} (\bibinfo {year} {2022})}\BibitemShut {NoStop}%
\bibitem [{\citenamefont {Lachance-Quirion}\ \emph {et~al.}(2024)\citenamefont {Lachance-Quirion}, \citenamefont {Lemonde}, \citenamefont {Simoneau}, \citenamefont {St-Jean}, \citenamefont {Lemieux}, \citenamefont {Turcotte}, \citenamefont {Wright}, \citenamefont {Lacroix}, \citenamefont {Fréchette-Viens}, \citenamefont {Shillito}, \citenamefont {Hopfmueller}, \citenamefont {Tremblay}, \citenamefont {Frattini}, \citenamefont {Camirand~Lemyre},\ and\ \citenamefont {St-Jean}}]{Lachance2023}%
  \BibitemOpen
  \bibfield  {author} {\bibinfo {author} {\bibfnamefont {D.}~\bibnamefont {Lachance-Quirion}}, \bibinfo {author} {\bibfnamefont {M.-A.}\ \bibnamefont {Lemonde}}, \bibinfo {author} {\bibfnamefont {J.~O.}\ \bibnamefont {Simoneau}}, \bibinfo {author} {\bibfnamefont {L.}~\bibnamefont {St-Jean}}, \bibinfo {author} {\bibfnamefont {P.}~\bibnamefont {Lemieux}}, \bibinfo {author} {\bibfnamefont {S.}~\bibnamefont {Turcotte}}, \bibinfo {author} {\bibfnamefont {W.}~\bibnamefont {Wright}}, \bibinfo {author} {\bibfnamefont {A.}~\bibnamefont {Lacroix}}, \bibinfo {author} {\bibfnamefont {J.}~\bibnamefont {Fréchette-Viens}}, \bibinfo {author} {\bibfnamefont {R.}~\bibnamefont {Shillito}}, \bibinfo {author} {\bibfnamefont {F.}~\bibnamefont {Hopfmueller}}, \bibinfo {author} {\bibfnamefont {M.}~\bibnamefont {Tremblay}}, \bibinfo {author} {\bibfnamefont {N.~E.}\ \bibnamefont {Frattini}}, \bibinfo {author} {\bibfnamefont {J.}~\bibnamefont {Camirand~Lemyre}},\ and\ \bibinfo {author} {\bibfnamefont {P.}~\bibnamefont {St-Jean}},\
  }\bibfield  {title} {\bibinfo {title} {Autonomous quantum error correction of gottesman-kitaev-preskill states},\ }\href {https://doi.org/10.1103/physrevlett.132.150607} {\bibfield  {journal} {\bibinfo  {journal} {Phys. Rev. Lett.}\ }\textbf {\bibinfo {volume} {132}},\ \bibinfo {pages} {15} (\bibinfo {year} {2024})}\BibitemShut {NoStop}%
\bibitem [{\citenamefont {Konno}\ \emph {et~al.}(2024)\citenamefont {Konno}, \citenamefont {Asavanant}, \citenamefont {Hanamura}, \citenamefont {Nagayoshi}, \citenamefont {Fukui}, \citenamefont {Sakaguchi}, \citenamefont {Ide}, \citenamefont {China}, \citenamefont {Yabuno}, \citenamefont {Miki}, \citenamefont {Terai}, \citenamefont {Takase}, \citenamefont {Endo}, \citenamefont {Marek}, \citenamefont {Filip}, \citenamefont {Loock},\ and\ \citenamefont {Furusawa}}]{konno2024}%
  \BibitemOpen
  \bibfield  {author} {\bibinfo {author} {\bibfnamefont {S.}~\bibnamefont {Konno}}, \bibinfo {author} {\bibfnamefont {W.}~\bibnamefont {Asavanant}}, \bibinfo {author} {\bibfnamefont {F.}~\bibnamefont {Hanamura}}, \bibinfo {author} {\bibfnamefont {H.}~\bibnamefont {Nagayoshi}}, \bibinfo {author} {\bibfnamefont {K.}~\bibnamefont {Fukui}}, \bibinfo {author} {\bibfnamefont {A.}~\bibnamefont {Sakaguchi}}, \bibinfo {author} {\bibfnamefont {R.}~\bibnamefont {Ide}}, \bibinfo {author} {\bibfnamefont {F.}~\bibnamefont {China}}, \bibinfo {author} {\bibfnamefont {M.}~\bibnamefont {Yabuno}}, \bibinfo {author} {\bibfnamefont {S.}~\bibnamefont {Miki}}, \bibinfo {author} {\bibfnamefont {H.}~\bibnamefont {Terai}}, \bibinfo {author} {\bibfnamefont {K.}~\bibnamefont {Takase}}, \bibinfo {author} {\bibfnamefont {M.}~\bibnamefont {Endo}}, \bibinfo {author} {\bibfnamefont {P.}~\bibnamefont {Marek}}, \bibinfo {author} {\bibfnamefont {R.}~\bibnamefont {Filip}}, \bibinfo {author} {\bibfnamefont {P.~v.}\ \bibnamefont {Loock}},\ and\
  \bibinfo {author} {\bibfnamefont {A.}~\bibnamefont {Furusawa}},\ }\bibfield  {title} {\bibinfo {title} {{Logical states for fault-tolerant quantum computation with propagating light}},\ }\href {https://doi.org/10.1126/science.adk7560} {\bibfield  {journal} {\bibinfo  {journal} {Science}\ }\textbf {\bibinfo {volume} {383}},\ \bibinfo {pages} {289} (\bibinfo {year} {2024})}\BibitemShut {NoStop}%
\bibitem [{\citenamefont {Fl\"{u}hmann}\ \emph {et~al.}(2019)\citenamefont {Fl\"{u}hmann}, \citenamefont {Nguyen}, \citenamefont {Marinelli}, \citenamefont {Negnevitsky}, \citenamefont {Mehta},\ and\ \citenamefont {Home}}]{Flhmann2019}%
  \BibitemOpen
  \bibfield  {author} {\bibinfo {author} {\bibfnamefont {C.}~\bibnamefont {Fl\"{u}hmann}}, \bibinfo {author} {\bibfnamefont {T.~L.}\ \bibnamefont {Nguyen}}, \bibinfo {author} {\bibfnamefont {M.}~\bibnamefont {Marinelli}}, \bibinfo {author} {\bibfnamefont {V.}~\bibnamefont {Negnevitsky}}, \bibinfo {author} {\bibfnamefont {K.}~\bibnamefont {Mehta}},\ and\ \bibinfo {author} {\bibfnamefont {J.~P.}\ \bibnamefont {Home}},\ }\bibfield  {title} {\bibinfo {title} {Encoding a qubit in a trapped-ion mechanical oscillator},\ }\href {https://doi.org/10.1038/s41586-019-0960-6} {\bibfield  {journal} {\bibinfo  {journal} {Nature}\ }\textbf {\bibinfo {volume} {566}},\ \bibinfo {pages} {513} (\bibinfo {year} {2019})}\BibitemShut {NoStop}%
\bibitem [{\citenamefont {Neeve}\ \emph {et~al.}(2022)\citenamefont {Neeve}, \citenamefont {Nguyen}, \citenamefont {Behrle},\ and\ \citenamefont {Home}}]{Neeve2022}%
  \BibitemOpen
  \bibfield  {author} {\bibinfo {author} {\bibfnamefont {B.~d.}\ \bibnamefont {Neeve}}, \bibinfo {author} {\bibfnamefont {T.-L.}\ \bibnamefont {Nguyen}}, \bibinfo {author} {\bibfnamefont {T.}~\bibnamefont {Behrle}},\ and\ \bibinfo {author} {\bibfnamefont {J.~P.}\ \bibnamefont {Home}},\ }\bibfield  {title} {\bibinfo {title} {{Error correction of a logical grid state qubit by dissipative pumping}},\ }\href {https://doi.org/10.1038/s41567-021-01487-7} {\bibfield  {journal} {\bibinfo  {journal} {Nat. Phys.}\ }\textbf {\bibinfo {volume} {18}},\ \bibinfo {pages} {296} (\bibinfo {year} {2022})}\BibitemShut {NoStop}%
\bibitem [{\citenamefont {Matsos}\ \emph {et~al.}(2024)\citenamefont {Matsos}, \citenamefont {Valahu}, \citenamefont {Navickas}, \citenamefont {Rao}, \citenamefont {Millican}, \citenamefont {Kolesnikow}, \citenamefont {Biercuk},\ and\ \citenamefont {Tan}}]{Matsos2024}%
  \BibitemOpen
  \bibfield  {author} {\bibinfo {author} {\bibfnamefont {V.~G.}\ \bibnamefont {Matsos}}, \bibinfo {author} {\bibfnamefont {C.~H.}\ \bibnamefont {Valahu}}, \bibinfo {author} {\bibfnamefont {T.}~\bibnamefont {Navickas}}, \bibinfo {author} {\bibfnamefont {A.~D.}\ \bibnamefont {Rao}}, \bibinfo {author} {\bibfnamefont {M.~J.}\ \bibnamefont {Millican}}, \bibinfo {author} {\bibfnamefont {X.~C.}\ \bibnamefont {Kolesnikow}}, \bibinfo {author} {\bibfnamefont {M.~J.}\ \bibnamefont {Biercuk}},\ and\ \bibinfo {author} {\bibfnamefont {T.~R.}\ \bibnamefont {Tan}},\ }\bibfield  {title} {\bibinfo {title} {Robust and deterministic preparation of bosonic logical states in a trapped ion},\ }\href {https://doi.org/10.1103/physrevlett.133.050602} {\bibfield  {journal} {\bibinfo  {journal} {Phys. Rev. Lett.}\ }\textbf {\bibinfo {volume} {133}},\ \bibinfo {pages} {5} (\bibinfo {year} {2024})}\BibitemShut {NoStop}%
\bibitem [{\citenamefont {Royer}\ \emph {et~al.}(2020)\citenamefont {Royer}, \citenamefont {Singh},\ and\ \citenamefont {Girvin}}]{Royer2020}%
  \BibitemOpen
  \bibfield  {author} {\bibinfo {author} {\bibfnamefont {B.}~\bibnamefont {Royer}}, \bibinfo {author} {\bibfnamefont {S.}~\bibnamefont {Singh}},\ and\ \bibinfo {author} {\bibfnamefont {S.}~\bibnamefont {Girvin}},\ }\bibfield  {title} {\bibinfo {title} {Stabilization of finite-energy gottesman-kitaev-preskill states},\ }\href {https://doi.org/10.1103/physrevlett.125.260509} {\bibfield  {journal} {\bibinfo  {journal} {Phys. Rev. Lett.}\ }\textbf {\bibinfo {volume} {125}},\ \bibinfo {pages} {26} (\bibinfo {year} {2020})}\BibitemShut {NoStop}%
\bibitem [{\citenamefont {Hastrup}\ and\ \citenamefont {Andersen}(2021)}]{Hastrup2021ir}%
  \BibitemOpen
  \bibfield  {author} {\bibinfo {author} {\bibfnamefont {J.}~\bibnamefont {Hastrup}}\ and\ \bibinfo {author} {\bibfnamefont {U.~L.}\ \bibnamefont {Andersen}},\ }\bibfield  {title} {\bibinfo {title} {Improved readout of qubit-coupled gottesman–kitaev–preskill states},\ }\href {https://doi.org/10.1088/2058-9565/ac070d} {\bibfield  {journal} {\bibinfo  {journal} {Quantum Sci. Technol.}\ }\textbf {\bibinfo {volume} {6}},\ \bibinfo {pages} {035016} (\bibinfo {year} {2021})}\BibitemShut {NoStop}%
\bibitem [{\citenamefont {Rojkov}\ \emph {et~al.}(2023)\citenamefont {Rojkov}, \citenamefont {R\"{o}ggla}, \citenamefont {Wagener}, \citenamefont {Fontboté-Schmidt}, \citenamefont {Welte}, \citenamefont {Home},\ and\ \citenamefont {Reiter}}]{Rojkov2023}%
  \BibitemOpen
  \bibfield  {author} {\bibinfo {author} {\bibfnamefont {I.}~\bibnamefont {Rojkov}}, \bibinfo {author} {\bibfnamefont {P.~M.}\ \bibnamefont {R\"{o}ggla}}, \bibinfo {author} {\bibfnamefont {M.}~\bibnamefont {Wagener}}, \bibinfo {author} {\bibfnamefont {M.}~\bibnamefont {Fontboté-Schmidt}}, \bibinfo {author} {\bibfnamefont {S.}~\bibnamefont {Welte}}, \bibinfo {author} {\bibfnamefont {J.}~\bibnamefont {Home}},\ and\ \bibinfo {author} {\bibfnamefont {F.}~\bibnamefont {Reiter}},\ }\href@noop {} {\bibinfo {title} {Two-qubit operations for finite-energy gottesman-kitaev-preskill encodings}} (\bibinfo {year} {2023}),\ \Eprint {https://arxiv.org/abs/2305.05262} {arXiv:2305.05262 [quant-ph]} \BibitemShut {NoStop}%
\bibitem [{\citenamefont {Ball}\ \emph {et~al.}(2021)\citenamefont {Ball}, \citenamefont {Biercuk}, \citenamefont {Carvalho}, \citenamefont {Chen}, \citenamefont {Hush}, \citenamefont {Castro}, \citenamefont {Li}, \citenamefont {Liebermann}, \citenamefont {Slatyer}, \citenamefont {Edmunds}, \citenamefont {Frey}, \citenamefont {Hempel},\ and\ \citenamefont {Milne}}]{boulder_opal1}%
  \BibitemOpen
  \bibfield  {author} {\bibinfo {author} {\bibfnamefont {H.}~\bibnamefont {Ball}}, \bibinfo {author} {\bibfnamefont {M.~J.}\ \bibnamefont {Biercuk}}, \bibinfo {author} {\bibfnamefont {A.~R.~R.}\ \bibnamefont {Carvalho}}, \bibinfo {author} {\bibfnamefont {J.}~\bibnamefont {Chen}}, \bibinfo {author} {\bibfnamefont {M.}~\bibnamefont {Hush}}, \bibinfo {author} {\bibfnamefont {L.~A.~D.}\ \bibnamefont {Castro}}, \bibinfo {author} {\bibfnamefont {L.}~\bibnamefont {Li}}, \bibinfo {author} {\bibfnamefont {P.~J.}\ \bibnamefont {Liebermann}}, \bibinfo {author} {\bibfnamefont {H.~J.}\ \bibnamefont {Slatyer}}, \bibinfo {author} {\bibfnamefont {C.}~\bibnamefont {Edmunds}}, \bibinfo {author} {\bibfnamefont {V.}~\bibnamefont {Frey}}, \bibinfo {author} {\bibfnamefont {C.}~\bibnamefont {Hempel}},\ and\ \bibinfo {author} {\bibfnamefont {A.}~\bibnamefont {Milne}},\ }\bibfield  {title} {\bibinfo {title} {Software tools for quantum control: improving quantum computer performance through noise and error suppression},\ }\href
  {https://doi.org/10.1088/2058-9565/abdca6} {\bibfield  {journal} {\bibinfo  {journal} {Quantum Sci. Technol.}\ }\textbf {\bibinfo {volume} {6}},\ \bibinfo {pages} {044011} (\bibinfo {year} {2021})}\BibitemShut {NoStop}%
\bibitem [{\citenamefont {Barenco}\ \emph {et~al.}(1995)\citenamefont {Barenco}, \citenamefont {Bennett}, \citenamefont {Cleve}, \citenamefont {DiVincenzo}, \citenamefont {Margolus}, \citenamefont {Shor}, \citenamefont {Sleator}, \citenamefont {Smolin},\ and\ \citenamefont {Weinfurter}}]{Barenco1995}%
  \BibitemOpen
  \bibfield  {author} {\bibinfo {author} {\bibfnamefont {A.}~\bibnamefont {Barenco}}, \bibinfo {author} {\bibfnamefont {C.~H.}\ \bibnamefont {Bennett}}, \bibinfo {author} {\bibfnamefont {R.}~\bibnamefont {Cleve}}, \bibinfo {author} {\bibfnamefont {D.~P.}\ \bibnamefont {DiVincenzo}}, \bibinfo {author} {\bibfnamefont {N.}~\bibnamefont {Margolus}}, \bibinfo {author} {\bibfnamefont {P.}~\bibnamefont {Shor}}, \bibinfo {author} {\bibfnamefont {T.}~\bibnamefont {Sleator}}, \bibinfo {author} {\bibfnamefont {J.~A.}\ \bibnamefont {Smolin}},\ and\ \bibinfo {author} {\bibfnamefont {H.}~\bibnamefont {Weinfurter}},\ }\bibfield  {title} {\bibinfo {title} {Elementary gates for quantum computation},\ }\href {https://doi.org/10.1103/physreva.52.3457} {\bibfield  {journal} {\bibinfo  {journal} {Phys. Rev. A}\ }\textbf {\bibinfo {volume} {52}},\ \bibinfo {pages} {3457–3467} (\bibinfo {year} {1995})}\BibitemShut {NoStop}%
\bibitem [{\citenamefont {Bremner}\ \emph {et~al.}(2002)\citenamefont {Bremner}, \citenamefont {Dawson}, \citenamefont {Dodd}, \citenamefont {Gilchrist}, \citenamefont {Harrow}, \citenamefont {Mortimer}, \citenamefont {Nielsen},\ and\ \citenamefont {Osborne}}]{Bremner2002}%
  \BibitemOpen
  \bibfield  {author} {\bibinfo {author} {\bibfnamefont {M.~J.}\ \bibnamefont {Bremner}}, \bibinfo {author} {\bibfnamefont {C.~M.}\ \bibnamefont {Dawson}}, \bibinfo {author} {\bibfnamefont {J.~L.}\ \bibnamefont {Dodd}}, \bibinfo {author} {\bibfnamefont {A.}~\bibnamefont {Gilchrist}}, \bibinfo {author} {\bibfnamefont {A.~W.}\ \bibnamefont {Harrow}}, \bibinfo {author} {\bibfnamefont {D.}~\bibnamefont {Mortimer}}, \bibinfo {author} {\bibfnamefont {M.~A.}\ \bibnamefont {Nielsen}},\ and\ \bibinfo {author} {\bibfnamefont {T.~J.}\ \bibnamefont {Osborne}},\ }\bibfield  {title} {\bibinfo {title} {Practical scheme for quantum computation with any two-qubit entangling gate},\ }\href {https://doi.org/10.1103/physrevlett.89.247902} {\bibfield  {journal} {\bibinfo  {journal} {Phys. Rev. Lett.}\ }\textbf {\bibinfo {volume} {89}},\ \bibinfo {pages} {24} (\bibinfo {year} {2002})}\BibitemShut {NoStop}%
\bibitem [{\citenamefont {Raussendorf}\ and\ \citenamefont {Briegel}(2001)}]{Raussendorf2001}%
  \BibitemOpen
  \bibfield  {author} {\bibinfo {author} {\bibfnamefont {R.}~\bibnamefont {Raussendorf}}\ and\ \bibinfo {author} {\bibfnamefont {H.~J.}\ \bibnamefont {Briegel}},\ }\bibfield  {title} {\bibinfo {title} {A one-way quantum computer},\ }\href {https://doi.org/10.1103/PhysRevLett.86.5188} {\bibfield  {journal} {\bibinfo  {journal} {Phys. Rev. Lett.}\ }\textbf {\bibinfo {volume} {86}},\ \bibinfo {pages} {5188} (\bibinfo {year} {2001})}\BibitemShut {NoStop}%
\bibitem [{\citenamefont {Valahu}\ \emph {et~al.}(2023)\citenamefont {Valahu}, \citenamefont {Olaya-Agudelo}, \citenamefont {MacDonell}, \citenamefont {Navickas}, \citenamefont {Rao}, \citenamefont {Millican}, \citenamefont {Pérez-Sánchez}, \citenamefont {Yuen-Zhou}, \citenamefont {Biercuk}, \citenamefont {Hempel}, \citenamefont {Tan},\ and\ \citenamefont {Kassal}}]{Valahu2023}%
  \BibitemOpen
  \bibfield  {author} {\bibinfo {author} {\bibfnamefont {C.~H.}\ \bibnamefont {Valahu}}, \bibinfo {author} {\bibfnamefont {V.~C.}\ \bibnamefont {Olaya-Agudelo}}, \bibinfo {author} {\bibfnamefont {R.~J.}\ \bibnamefont {MacDonell}}, \bibinfo {author} {\bibfnamefont {T.}~\bibnamefont {Navickas}}, \bibinfo {author} {\bibfnamefont {A.~D.}\ \bibnamefont {Rao}}, \bibinfo {author} {\bibfnamefont {M.~J.}\ \bibnamefont {Millican}}, \bibinfo {author} {\bibfnamefont {J.~B.}\ \bibnamefont {Pérez-Sánchez}}, \bibinfo {author} {\bibfnamefont {J.}~\bibnamefont {Yuen-Zhou}}, \bibinfo {author} {\bibfnamefont {M.~J.}\ \bibnamefont {Biercuk}}, \bibinfo {author} {\bibfnamefont {C.}~\bibnamefont {Hempel}}, \bibinfo {author} {\bibfnamefont {T.~R.}\ \bibnamefont {Tan}},\ and\ \bibinfo {author} {\bibfnamefont {I.}~\bibnamefont {Kassal}},\ }\bibfield  {title} {\bibinfo {title} {Direct observation of geometric-phase interference in dynamics around a conical intersection},\ }\href {https://doi.org/10.1038/s41557-023-01300-3} {\bibfield
   {journal} {\bibinfo  {journal} {Nat. Chem.}\ }\textbf {\bibinfo {volume} {15}},\ \bibinfo {pages} {1503–1508} (\bibinfo {year} {2023})}\BibitemShut {NoStop}%
\bibitem [{\citenamefont {Tan}\ \emph {et~al.}(2023)\citenamefont {Tan}, \citenamefont {Navickas}, \citenamefont {Valahu}, \citenamefont {Jee}, \citenamefont {Rao}, \citenamefont {Millican},\ and\ \citenamefont {Biercuk}}]{Tan2023}%
  \BibitemOpen
  \bibfield  {author} {\bibinfo {author} {\bibfnamefont {T.~R.}\ \bibnamefont {Tan}}, \bibinfo {author} {\bibfnamefont {T.}~\bibnamefont {Navickas}}, \bibinfo {author} {\bibfnamefont {C.}~\bibnamefont {Valahu}}, \bibinfo {author} {\bibfnamefont {J.}~\bibnamefont {Jee}}, \bibinfo {author} {\bibfnamefont {A.}~\bibnamefont {Rao}}, \bibinfo {author} {\bibfnamefont {M.}~\bibnamefont {Millican}},\ and\ \bibinfo {author} {\bibfnamefont {M.}~\bibnamefont {Biercuk}},\ }\bibfield  {title} {\bibinfo {title} {Improving a trapped-ion quantum computer with a cryogenic sapphire oscillator},\ }\href {https://doi.org/10.1109/EFTF/IFCS57587.2023.10272197} {\bibfield  {journal} {\bibinfo  {journal} {2023 Joint Conference of the European Frequency and Time Forum and IEEE International Frequency Control Symposium (EFTF/IFCS)}\ ,\ \bibinfo {pages} {1}} (\bibinfo {year} {2023})}\BibitemShut {NoStop}%
\bibitem [{\citenamefont {Wineland}\ \emph {et~al.}(1998)\citenamefont {Wineland}, \citenamefont {Monroe}, \citenamefont {Itano}, \citenamefont {Leibfried}, \citenamefont {King},\ and\ \citenamefont {Meekhof}}]{Wineland1998}%
  \BibitemOpen
  \bibfield  {author} {\bibinfo {author} {\bibfnamefont {D.~J.}\ \bibnamefont {Wineland}}, \bibinfo {author} {\bibfnamefont {C.}~\bibnamefont {Monroe}}, \bibinfo {author} {\bibfnamefont {W.~M.}\ \bibnamefont {Itano}}, \bibinfo {author} {\bibfnamefont {D.}~\bibnamefont {Leibfried}}, \bibinfo {author} {\bibfnamefont {B.~E.}\ \bibnamefont {King}},\ and\ \bibinfo {author} {\bibfnamefont {D.~M.}\ \bibnamefont {Meekhof}},\ }\bibfield  {title} {\bibinfo {title} {{Experimental issues in coherent quantum-state manipulation of trapped atomic ions}},\ }\href {https://doi.org/10.6028/jres.103.019} {\bibfield  {journal} {\bibinfo  {journal} {J. Res. Natl. Inst. Stan.}\ }\textbf {\bibinfo {volume} {103}},\ \bibinfo {pages} {259} (\bibinfo {year} {1998})}\BibitemShut {NoStop}%
\bibitem [{\citenamefont {Shaw}\ \emph {et~al.}(2024{\natexlab{a}})\citenamefont {Shaw}, \citenamefont {Doherty},\ and\ \citenamefont {Grimsmo}}]{Shaw2024}%
  \BibitemOpen
  \bibfield  {author} {\bibinfo {author} {\bibfnamefont {M.~H.}\ \bibnamefont {Shaw}}, \bibinfo {author} {\bibfnamefont {A.~C.}\ \bibnamefont {Doherty}},\ and\ \bibinfo {author} {\bibfnamefont {A.~L.}\ \bibnamefont {Grimsmo}},\ }\bibfield  {title} {\bibinfo {title} {Stabilizer subsystem decompositions for single- and multimode gottesman-kitaev-preskill codes},\ }\href {https://doi.org/10.1103/prxquantum.5.010331} {\bibfield  {journal} {\bibinfo  {journal} {PRX Quantum}\ }\textbf {\bibinfo {volume} {5}},\ \bibinfo {pages} {010331} (\bibinfo {year} {2024}{\natexlab{a}})}\BibitemShut {NoStop}%
\bibitem [{\citenamefont {Shaw}\ \emph {et~al.}(2024{\natexlab{b}})\citenamefont {Shaw}, \citenamefont {Doherty},\ and\ \citenamefont {Grimsmo}}]{Shaw2024a}%
  \BibitemOpen
  \bibfield  {author} {\bibinfo {author} {\bibfnamefont {M.~H.}\ \bibnamefont {Shaw}}, \bibinfo {author} {\bibfnamefont {A.~C.}\ \bibnamefont {Doherty}},\ and\ \bibinfo {author} {\bibfnamefont {A.~L.}\ \bibnamefont {Grimsmo}},\ }\href@noop {} {\bibinfo {title} {Logical gates and read-out of superconducting gottesman-kitaev-preskill qubits}} (\bibinfo {year} {2024}{\natexlab{b}}),\ \Eprint {https://arxiv.org/abs/2403.02396} {arXiv:2403.02396 [quant-ph]} \BibitemShut {NoStop}%
\bibitem [{\citenamefont {Bourassa}\ \emph {et~al.}(2021)\citenamefont {Bourassa}, \citenamefont {Alexander}, \citenamefont {Vasmer}, \citenamefont {Patil}, \citenamefont {Tzitrin}, \citenamefont {Matsuura}, \citenamefont {Su}, \citenamefont {Baragiola}, \citenamefont {Guha}, \citenamefont {Dauphinais}, \citenamefont {Sabapathy}, \citenamefont {Menicucci},\ and\ \citenamefont {Dhand}}]{Bourassa2021}%
  \BibitemOpen
  \bibfield  {author} {\bibinfo {author} {\bibfnamefont {J.~E.}\ \bibnamefont {Bourassa}}, \bibinfo {author} {\bibfnamefont {R.~N.}\ \bibnamefont {Alexander}}, \bibinfo {author} {\bibfnamefont {M.}~\bibnamefont {Vasmer}}, \bibinfo {author} {\bibfnamefont {A.}~\bibnamefont {Patil}}, \bibinfo {author} {\bibfnamefont {I.}~\bibnamefont {Tzitrin}}, \bibinfo {author} {\bibfnamefont {T.}~\bibnamefont {Matsuura}}, \bibinfo {author} {\bibfnamefont {D.}~\bibnamefont {Su}}, \bibinfo {author} {\bibfnamefont {B.~Q.}\ \bibnamefont {Baragiola}}, \bibinfo {author} {\bibfnamefont {S.}~\bibnamefont {Guha}}, \bibinfo {author} {\bibfnamefont {G.}~\bibnamefont {Dauphinais}}, \bibinfo {author} {\bibfnamefont {K.~K.}\ \bibnamefont {Sabapathy}}, \bibinfo {author} {\bibfnamefont {N.~C.}\ \bibnamefont {Menicucci}},\ and\ \bibinfo {author} {\bibfnamefont {I.}~\bibnamefont {Dhand}},\ }\bibfield  {title} {\bibinfo {title} {Blueprint for a scalable photonic fault-tolerant quantum computer},\ }\href
  {https://doi.org/10.22331/q-2021-02-04-392} {\bibfield  {journal} {\bibinfo  {journal} {Quantum}\ }\textbf {\bibinfo {volume} {5}},\ \bibinfo {pages} {392} (\bibinfo {year} {2021})}\BibitemShut {NoStop}%
\bibitem [{\citenamefont {Kolesnikow}\ \emph {et~al.}(2024)\citenamefont {Kolesnikow}, \citenamefont {Bomantara}, \citenamefont {Doherty},\ and\ \citenamefont {Grimsmo}}]{Kolesnikow2024}%
  \BibitemOpen
  \bibfield  {author} {\bibinfo {author} {\bibfnamefont {X.~C.}\ \bibnamefont {Kolesnikow}}, \bibinfo {author} {\bibfnamefont {R.~W.}\ \bibnamefont {Bomantara}}, \bibinfo {author} {\bibfnamefont {A.~C.}\ \bibnamefont {Doherty}},\ and\ \bibinfo {author} {\bibfnamefont {A.~L.}\ \bibnamefont {Grimsmo}},\ }\bibfield  {title} {\bibinfo {title} {Gottesman-kitaev-preskill state preparation using periodic driving},\ }\href {https://doi.org/10.1103/PhysRevLett.132.130605} {\bibfield  {journal} {\bibinfo  {journal} {Phys. Rev. Lett.}\ }\textbf {\bibinfo {volume} {132}},\ \bibinfo {pages} {130605} (\bibinfo {year} {2024})}\BibitemShut {NoStop}%
\bibitem [{\citenamefont {Sch\"{a}fer}\ \emph {et~al.}(2018)\citenamefont {Sch\"{a}fer}, \citenamefont {Ballance}, \citenamefont {Thirumalai}, \citenamefont {Stephenson}, \citenamefont {Ballance}, \citenamefont {Steane},\ and\ \citenamefont {Lucas}}]{Schafer2018}%
  \BibitemOpen
  \bibfield  {author} {\bibinfo {author} {\bibfnamefont {V.~M.}\ \bibnamefont {Sch\"{a}fer}}, \bibinfo {author} {\bibfnamefont {C.~J.}\ \bibnamefont {Ballance}}, \bibinfo {author} {\bibfnamefont {K.}~\bibnamefont {Thirumalai}}, \bibinfo {author} {\bibfnamefont {L.~J.}\ \bibnamefont {Stephenson}}, \bibinfo {author} {\bibfnamefont {T.~G.}\ \bibnamefont {Ballance}}, \bibinfo {author} {\bibfnamefont {A.~M.}\ \bibnamefont {Steane}},\ and\ \bibinfo {author} {\bibfnamefont {D.~M.}\ \bibnamefont {Lucas}},\ }\bibfield  {title} {\bibinfo {title} {Fast quantum logic gates with trapped-ion qubits},\ }\href {https://doi.org/10.1038/nature25737} {\bibfield  {journal} {\bibinfo  {journal} {Nature}\ }\textbf {\bibinfo {volume} {555}},\ \bibinfo {pages} {75–78} (\bibinfo {year} {2018})}\BibitemShut {NoStop}%
\bibitem [{\citenamefont {Chou}\ \emph {et~al.}(2017)\citenamefont {Chou}, \citenamefont {Kurz}, \citenamefont {Hume}, \citenamefont {Plessow}, \citenamefont {Leibrandt},\ and\ \citenamefont {Leibfried}}]{Chou2017}%
  \BibitemOpen
  \bibfield  {author} {\bibinfo {author} {\bibfnamefont {C.-W.}\ \bibnamefont {Chou}}, \bibinfo {author} {\bibfnamefont {C.}~\bibnamefont {Kurz}}, \bibinfo {author} {\bibfnamefont {D.~B.}\ \bibnamefont {Hume}}, \bibinfo {author} {\bibfnamefont {P.~N.}\ \bibnamefont {Plessow}}, \bibinfo {author} {\bibfnamefont {D.~R.}\ \bibnamefont {Leibrandt}},\ and\ \bibinfo {author} {\bibfnamefont {D.}~\bibnamefont {Leibfried}},\ }\bibfield  {title} {\bibinfo {title} {Preparation and coherent manipulation of pure quantum states of a single molecular ion},\ }\href {https://doi.org/10.1038/nature22338} {\bibfield  {journal} {\bibinfo  {journal} {Nature}\ }\textbf {\bibinfo {volume} {545}},\ \bibinfo {pages} {203–207} (\bibinfo {year} {2017})}\BibitemShut {NoStop}%
\bibitem [{\citenamefont {Postler}\ \emph {et~al.}(2022)\citenamefont {Postler}, \citenamefont {Heu{\textbeta}en}, \citenamefont {Pogorelov}, \citenamefont {Rispler}, \citenamefont {Feldker}, \citenamefont {Meth}, \citenamefont {Marciniak}, \citenamefont {Stricker}, \citenamefont {Ringbauer}, \citenamefont {Blatt}, \citenamefont {Schindler}, \citenamefont {Müller},\ and\ \citenamefont {Monz}}]{Postler2022}%
  \BibitemOpen
  \bibfield  {author} {\bibinfo {author} {\bibfnamefont {L.}~\bibnamefont {Postler}}, \bibinfo {author} {\bibfnamefont {S.}~\bibnamefont {Heu{\textbeta}en}}, \bibinfo {author} {\bibfnamefont {I.}~\bibnamefont {Pogorelov}}, \bibinfo {author} {\bibfnamefont {M.}~\bibnamefont {Rispler}}, \bibinfo {author} {\bibfnamefont {T.}~\bibnamefont {Feldker}}, \bibinfo {author} {\bibfnamefont {M.}~\bibnamefont {Meth}}, \bibinfo {author} {\bibfnamefont {C.~D.}\ \bibnamefont {Marciniak}}, \bibinfo {author} {\bibfnamefont {R.}~\bibnamefont {Stricker}}, \bibinfo {author} {\bibfnamefont {M.}~\bibnamefont {Ringbauer}}, \bibinfo {author} {\bibfnamefont {R.}~\bibnamefont {Blatt}}, \bibinfo {author} {\bibfnamefont {P.}~\bibnamefont {Schindler}}, \bibinfo {author} {\bibfnamefont {M.}~\bibnamefont {Müller}},\ and\ \bibinfo {author} {\bibfnamefont {T.}~\bibnamefont {Monz}},\ }\bibfield  {title} {\bibinfo {title} {Demonstration of fault-tolerant universal quantum gate operations},\ }\href {https://doi.org/10.1038/s41586-022-04721-1}
  {\bibfield  {journal} {\bibinfo  {journal} {Nature}\ }\textbf {\bibinfo {volume} {605}},\ \bibinfo {pages} {675} (\bibinfo {year} {2022})},\ \Eprint {https://arxiv.org/abs/2111.12654} {2111.12654} \BibitemShut {NoStop}%
\bibitem [{\citenamefont {Chen}\ \emph {et~al.}(2023)\citenamefont {Chen}, \citenamefont {Nielsen}, \citenamefont {Ebert}, \citenamefont {Inlek}, \citenamefont {Wright}, \citenamefont {Chaplin}, \citenamefont {Maksymov}, \citenamefont {Páez}, \citenamefont {Poudel}, \citenamefont {Maunz},\ and\ \citenamefont {Gamble}}]{chen2023arXiv}%
  \BibitemOpen
  \bibfield  {author} {\bibinfo {author} {\bibfnamefont {J.-S.}\ \bibnamefont {Chen}}, \bibinfo {author} {\bibfnamefont {E.}~\bibnamefont {Nielsen}}, \bibinfo {author} {\bibfnamefont {M.}~\bibnamefont {Ebert}}, \bibinfo {author} {\bibfnamefont {V.}~\bibnamefont {Inlek}}, \bibinfo {author} {\bibfnamefont {K.}~\bibnamefont {Wright}}, \bibinfo {author} {\bibfnamefont {V.}~\bibnamefont {Chaplin}}, \bibinfo {author} {\bibfnamefont {A.}~\bibnamefont {Maksymov}}, \bibinfo {author} {\bibfnamefont {E.}~\bibnamefont {Páez}}, \bibinfo {author} {\bibfnamefont {A.}~\bibnamefont {Poudel}}, \bibinfo {author} {\bibfnamefont {P.}~\bibnamefont {Maunz}},\ and\ \bibinfo {author} {\bibfnamefont {J.}~\bibnamefont {Gamble}},\ }\href {https://arxiv.org/abs/2308.05071} {\bibinfo {title} {Benchmarking a trapped-ion quantum computer with 29 algorithmic qubits}} (\bibinfo {year} {2023}),\ \Eprint {https://arxiv.org/abs/2308.05071} {arXiv:2308.05071 [quant-ph]} \BibitemShut {NoStop}%
\bibitem [{\citenamefont {Warring}\ \emph {et~al.}(2020)\citenamefont {Warring}, \citenamefont {Hakelberg}, \citenamefont {Kiefer}, \citenamefont {Wittemer},\ and\ \citenamefont {Schaetz}}]{Warring2020}%
  \BibitemOpen
  \bibfield  {author} {\bibinfo {author} {\bibfnamefont {U.}~\bibnamefont {Warring}}, \bibinfo {author} {\bibfnamefont {F.}~\bibnamefont {Hakelberg}}, \bibinfo {author} {\bibfnamefont {P.}~\bibnamefont {Kiefer}}, \bibinfo {author} {\bibfnamefont {M.}~\bibnamefont {Wittemer}},\ and\ \bibinfo {author} {\bibfnamefont {T.}~\bibnamefont {Schaetz}},\ }\bibfield  {title} {\bibinfo {title} {Trapped ion architecture for multi‐dimensional quantum simulations},\ }\href {https://doi.org/10.1002/qute.201900137} {\bibfield  {journal} {\bibinfo  {journal} {Adv. Quantum Technol.}\ }\textbf {\bibinfo {volume} {3}},\ \bibinfo {pages} {11} (\bibinfo {year} {2020})}\BibitemShut {NoStop}%
\bibitem [{\citenamefont {Kiesenhofer}\ \emph {et~al.}(2023)\citenamefont {Kiesenhofer}, \citenamefont {Hainzer}, \citenamefont {Zhdanov}, \citenamefont {Holz}, \citenamefont {Bock}, \citenamefont {Ollikainen},\ and\ \citenamefont {Roos}}]{Kiesenhofer2023}%
  \BibitemOpen
  \bibfield  {author} {\bibinfo {author} {\bibfnamefont {D.}~\bibnamefont {Kiesenhofer}}, \bibinfo {author} {\bibfnamefont {H.}~\bibnamefont {Hainzer}}, \bibinfo {author} {\bibfnamefont {A.}~\bibnamefont {Zhdanov}}, \bibinfo {author} {\bibfnamefont {P.~C.}\ \bibnamefont {Holz}}, \bibinfo {author} {\bibfnamefont {M.}~\bibnamefont {Bock}}, \bibinfo {author} {\bibfnamefont {T.}~\bibnamefont {Ollikainen}},\ and\ \bibinfo {author} {\bibfnamefont {C.~F.}\ \bibnamefont {Roos}},\ }\bibfield  {title} {\bibinfo {title} {Controlling two-dimensional coulomb crystals of more than 100 ions in a monolithic radio-frequency trap},\ }\href {https://doi.org/10.1103/PRXQuantum.4.020317} {\bibfield  {journal} {\bibinfo  {journal} {PRX Quantum}\ }\textbf {\bibinfo {volume} {4}},\ \bibinfo {pages} {020317} (\bibinfo {year} {2023})}\BibitemShut {NoStop}%
\bibitem [{\citenamefont {Guo}\ \emph {et~al.}(2024)\citenamefont {Guo}, \citenamefont {Wu}, \citenamefont {Ye}, \citenamefont {Zhang}, \citenamefont {Lian}, \citenamefont {Yao}, \citenamefont {Wang}, \citenamefont {Yan}, \citenamefont {Yi}, \citenamefont {Xu}, \citenamefont {Li}, \citenamefont {Hou}, \citenamefont {Xu}, \citenamefont {Guo}, \citenamefont {Zhang}, \citenamefont {Qi}, \citenamefont {Zhou}, \citenamefont {He},\ and\ \citenamefont {Duan}}]{Guo2024}%
  \BibitemOpen
  \bibfield  {author} {\bibinfo {author} {\bibfnamefont {S.-A.}\ \bibnamefont {Guo}}, \bibinfo {author} {\bibfnamefont {Y.-K.}\ \bibnamefont {Wu}}, \bibinfo {author} {\bibfnamefont {J.}~\bibnamefont {Ye}}, \bibinfo {author} {\bibfnamefont {L.}~\bibnamefont {Zhang}}, \bibinfo {author} {\bibfnamefont {W.-Q.}\ \bibnamefont {Lian}}, \bibinfo {author} {\bibfnamefont {R.}~\bibnamefont {Yao}}, \bibinfo {author} {\bibfnamefont {Y.}~\bibnamefont {Wang}}, \bibinfo {author} {\bibfnamefont {R.-Y.}\ \bibnamefont {Yan}}, \bibinfo {author} {\bibfnamefont {Y.-J.}\ \bibnamefont {Yi}}, \bibinfo {author} {\bibfnamefont {Y.-L.}\ \bibnamefont {Xu}}, \bibinfo {author} {\bibfnamefont {B.-W.}\ \bibnamefont {Li}}, \bibinfo {author} {\bibfnamefont {Y.-H.}\ \bibnamefont {Hou}}, \bibinfo {author} {\bibfnamefont {Y.-Z.}\ \bibnamefont {Xu}}, \bibinfo {author} {\bibfnamefont {W.-X.}\ \bibnamefont {Guo}}, \bibinfo {author} {\bibfnamefont {C.}~\bibnamefont {Zhang}}, \bibinfo {author} {\bibfnamefont {B.-X.}\ \bibnamefont {Qi}}, \bibinfo
  {author} {\bibfnamefont {Z.-C.}\ \bibnamefont {Zhou}}, \bibinfo {author} {\bibfnamefont {L.}~\bibnamefont {He}},\ and\ \bibinfo {author} {\bibfnamefont {L.-M.}\ \bibnamefont {Duan}},\ }\bibfield  {title} {\bibinfo {title} {A site-resolved two-dimensional quantum simulator with hundreds of trapped ions},\ }\href {https://doi.org/10.1038/s41586-024-07459-0} {\bibfield  {journal} {\bibinfo  {journal} {Nature}\ }\textbf {\bibinfo {volume} {630}},\ \bibinfo {pages} {613–618} (\bibinfo {year} {2024})}\BibitemShut {NoStop}%
\bibitem [{\citenamefont {Lysne}\ \emph {et~al.}(2024)\citenamefont {Lysne}, \citenamefont {Niedermeyer}, \citenamefont {Wilson}, \citenamefont {Slichter},\ and\ \citenamefont {Leibfried}}]{Lysne2024}%
  \BibitemOpen
  \bibfield  {author} {\bibinfo {author} {\bibfnamefont {N.~K.}\ \bibnamefont {Lysne}}, \bibinfo {author} {\bibfnamefont {J.~F.}\ \bibnamefont {Niedermeyer}}, \bibinfo {author} {\bibfnamefont {A.~C.}\ \bibnamefont {Wilson}}, \bibinfo {author} {\bibfnamefont {D.~H.}\ \bibnamefont {Slichter}},\ and\ \bibinfo {author} {\bibfnamefont {D.}~\bibnamefont {Leibfried}},\ }\bibfield  {title} {\bibinfo {title} {Individual addressing and state readout of trapped ions utilizing radio-frequency micromotion},\ }\href {https://doi.org/10.1103/physrevlett.133.033201} {\bibfield  {journal} {\bibinfo  {journal} {Phys. Rev. Lett.}\ }\textbf {\bibinfo {volume} {133}},\ \bibinfo {pages} {3} (\bibinfo {year} {2024})}\BibitemShut {NoStop}%
\bibitem [{\citenamefont {Moses}\ \emph {et~al.}(2023)\citenamefont {Moses}, \citenamefont {Baldwin}, \citenamefont {Allman}, \citenamefont {Ancona}, \citenamefont {Ascarrunz}, \citenamefont {Barnes}, \citenamefont {Bartolotta}, \citenamefont {Bjork}, \citenamefont {Blanchard}, \citenamefont {Bohn}, \citenamefont {Bohnet}, \citenamefont {Brown}, \citenamefont {Burdick}, \citenamefont {Burton}, \citenamefont {Campbell}, \citenamefont {Campora}, \citenamefont {Carron}, \citenamefont {Chambers}, \citenamefont {Chan}, \citenamefont {Chen}, \citenamefont {Chernoguzov}, \citenamefont {Chertkov}, \citenamefont {Colina}, \citenamefont {Curtis}, \citenamefont {Daniel}, \citenamefont {DeCross}, \citenamefont {Deen}, \citenamefont {Delaney}, \citenamefont {Dreiling}, \citenamefont {Ertsgaard}, \citenamefont {Esposito}, \citenamefont {Estey}, \citenamefont {Fabrikant}, \citenamefont {Figgatt}, \citenamefont {Foltz}, \citenamefont {Foss-Feig}, \citenamefont {Francois}, \citenamefont {Gaebler}, \citenamefont {Gatterman},
  \citenamefont {Gilbreth}, \citenamefont {Giles}, \citenamefont {Glynn}, \citenamefont {Hall}, \citenamefont {Hankin}, \citenamefont {Hansen}, \citenamefont {Hayes}, \citenamefont {Higashi}, \citenamefont {Hoffman}, \citenamefont {Horning}, \citenamefont {Hout}, \citenamefont {Jacobs}, \citenamefont {Johansen}, \citenamefont {Jones}, \citenamefont {Karcz}, \citenamefont {Klein}, \citenamefont {Lauria}, \citenamefont {Lee}, \citenamefont {Liefer}, \citenamefont {Lu}, \citenamefont {Lucchetti}, \citenamefont {Lytle}, \citenamefont {Malm}, \citenamefont {Matheny}, \citenamefont {Mathewson}, \citenamefont {Mayer}, \citenamefont {Miller}, \citenamefont {Mills}, \citenamefont {Neyenhuis}, \citenamefont {Nugent}, \citenamefont {Olson}, \citenamefont {Parks}, \citenamefont {Price}, \citenamefont {Price}, \citenamefont {Pugh}, \citenamefont {Ransford}, \citenamefont {Reed}, \citenamefont {Roman}, \citenamefont {Rowe}, \citenamefont {Ryan-Anderson}, \citenamefont {Sanders}, \citenamefont {Sedlacek}, \citenamefont
  {Shevchuk}, \citenamefont {Siegfried}, \citenamefont {Skripka}, \citenamefont {Spaun}, \citenamefont {Sprenkle}, \citenamefont {Stutz}, \citenamefont {Swallows}, \citenamefont {Tobey}, \citenamefont {Tran}, \citenamefont {Tran}, \citenamefont {Vogt}, \citenamefont {Volin}, \citenamefont {Walker}, \citenamefont {Zolot},\ and\ \citenamefont {Pino}}]{Moses2023}%
  \BibitemOpen
  \bibfield  {author} {\bibinfo {author} {\bibfnamefont {S.}~\bibnamefont {Moses}}, \bibinfo {author} {\bibfnamefont {C.}~\bibnamefont {Baldwin}}, \bibinfo {author} {\bibfnamefont {M.}~\bibnamefont {Allman}}, \bibinfo {author} {\bibfnamefont {R.}~\bibnamefont {Ancona}}, \bibinfo {author} {\bibfnamefont {L.}~\bibnamefont {Ascarrunz}}, \bibinfo {author} {\bibfnamefont {C.}~\bibnamefont {Barnes}}, \bibinfo {author} {\bibfnamefont {J.}~\bibnamefont {Bartolotta}}, \bibinfo {author} {\bibfnamefont {B.}~\bibnamefont {Bjork}}, \bibinfo {author} {\bibfnamefont {P.}~\bibnamefont {Blanchard}}, \bibinfo {author} {\bibfnamefont {M.}~\bibnamefont {Bohn}}, \bibinfo {author} {\bibfnamefont {J.}~\bibnamefont {Bohnet}}, \bibinfo {author} {\bibfnamefont {N.}~\bibnamefont {Brown}}, \bibinfo {author} {\bibfnamefont {N.}~\bibnamefont {Burdick}}, \bibinfo {author} {\bibfnamefont {W.}~\bibnamefont {Burton}}, \bibinfo {author} {\bibfnamefont {S.}~\bibnamefont {Campbell}}, \bibinfo {author} {\bibfnamefont {J.}~\bibnamefont {Campora}},
  \bibinfo {author} {\bibfnamefont {C.}~\bibnamefont {Carron}}, \bibinfo {author} {\bibfnamefont {J.}~\bibnamefont {Chambers}}, \bibinfo {author} {\bibfnamefont {J.}~\bibnamefont {Chan}}, \bibinfo {author} {\bibfnamefont {Y.}~\bibnamefont {Chen}}, \bibinfo {author} {\bibfnamefont {A.}~\bibnamefont {Chernoguzov}}, \bibinfo {author} {\bibfnamefont {E.}~\bibnamefont {Chertkov}}, \bibinfo {author} {\bibfnamefont {J.}~\bibnamefont {Colina}}, \bibinfo {author} {\bibfnamefont {J.}~\bibnamefont {Curtis}}, \bibinfo {author} {\bibfnamefont {R.}~\bibnamefont {Daniel}}, \bibinfo {author} {\bibfnamefont {M.}~\bibnamefont {DeCross}}, \bibinfo {author} {\bibfnamefont {D.}~\bibnamefont {Deen}}, \bibinfo {author} {\bibfnamefont {C.}~\bibnamefont {Delaney}}, \bibinfo {author} {\bibfnamefont {J.}~\bibnamefont {Dreiling}}, \bibinfo {author} {\bibfnamefont {C.}~\bibnamefont {Ertsgaard}}, \bibinfo {author} {\bibfnamefont {J.}~\bibnamefont {Esposito}}, \bibinfo {author} {\bibfnamefont {B.}~\bibnamefont {Estey}}, \bibinfo {author}
  {\bibfnamefont {M.}~\bibnamefont {Fabrikant}}, \bibinfo {author} {\bibfnamefont {C.}~\bibnamefont {Figgatt}}, \bibinfo {author} {\bibfnamefont {C.}~\bibnamefont {Foltz}}, \bibinfo {author} {\bibfnamefont {M.}~\bibnamefont {Foss-Feig}}, \bibinfo {author} {\bibfnamefont {D.}~\bibnamefont {Francois}}, \bibinfo {author} {\bibfnamefont {J.}~\bibnamefont {Gaebler}}, \bibinfo {author} {\bibfnamefont {T.}~\bibnamefont {Gatterman}}, \bibinfo {author} {\bibfnamefont {C.}~\bibnamefont {Gilbreth}}, \bibinfo {author} {\bibfnamefont {J.}~\bibnamefont {Giles}}, \bibinfo {author} {\bibfnamefont {E.}~\bibnamefont {Glynn}}, \bibinfo {author} {\bibfnamefont {A.}~\bibnamefont {Hall}}, \bibinfo {author} {\bibfnamefont {A.}~\bibnamefont {Hankin}}, \bibinfo {author} {\bibfnamefont {A.}~\bibnamefont {Hansen}}, \bibinfo {author} {\bibfnamefont {D.}~\bibnamefont {Hayes}}, \bibinfo {author} {\bibfnamefont {B.}~\bibnamefont {Higashi}}, \bibinfo {author} {\bibfnamefont {I.}~\bibnamefont {Hoffman}}, \bibinfo {author} {\bibfnamefont
  {B.}~\bibnamefont {Horning}}, \bibinfo {author} {\bibfnamefont {J.}~\bibnamefont {Hout}}, \bibinfo {author} {\bibfnamefont {R.}~\bibnamefont {Jacobs}}, \bibinfo {author} {\bibfnamefont {J.}~\bibnamefont {Johansen}}, \bibinfo {author} {\bibfnamefont {L.}~\bibnamefont {Jones}}, \bibinfo {author} {\bibfnamefont {J.}~\bibnamefont {Karcz}}, \bibinfo {author} {\bibfnamefont {T.}~\bibnamefont {Klein}}, \bibinfo {author} {\bibfnamefont {P.}~\bibnamefont {Lauria}}, \bibinfo {author} {\bibfnamefont {P.}~\bibnamefont {Lee}}, \bibinfo {author} {\bibfnamefont {D.}~\bibnamefont {Liefer}}, \bibinfo {author} {\bibfnamefont {S.}~\bibnamefont {Lu}}, \bibinfo {author} {\bibfnamefont {D.}~\bibnamefont {Lucchetti}}, \bibinfo {author} {\bibfnamefont {C.}~\bibnamefont {Lytle}}, \bibinfo {author} {\bibfnamefont {A.}~\bibnamefont {Malm}}, \bibinfo {author} {\bibfnamefont {M.}~\bibnamefont {Matheny}}, \bibinfo {author} {\bibfnamefont {B.}~\bibnamefont {Mathewson}}, \bibinfo {author} {\bibfnamefont {K.}~\bibnamefont {Mayer}},
  \bibinfo {author} {\bibfnamefont {D.}~\bibnamefont {Miller}}, \bibinfo {author} {\bibfnamefont {M.}~\bibnamefont {Mills}}, \bibinfo {author} {\bibfnamefont {B.}~\bibnamefont {Neyenhuis}}, \bibinfo {author} {\bibfnamefont {L.}~\bibnamefont {Nugent}}, \bibinfo {author} {\bibfnamefont {S.}~\bibnamefont {Olson}}, \bibinfo {author} {\bibfnamefont {J.}~\bibnamefont {Parks}}, \bibinfo {author} {\bibfnamefont {G.}~\bibnamefont {Price}}, \bibinfo {author} {\bibfnamefont {Z.}~\bibnamefont {Price}}, \bibinfo {author} {\bibfnamefont {M.}~\bibnamefont {Pugh}}, \bibinfo {author} {\bibfnamefont {A.}~\bibnamefont {Ransford}}, \bibinfo {author} {\bibfnamefont {A.}~\bibnamefont {Reed}}, \bibinfo {author} {\bibfnamefont {C.}~\bibnamefont {Roman}}, \bibinfo {author} {\bibfnamefont {M.}~\bibnamefont {Rowe}}, \bibinfo {author} {\bibfnamefont {C.}~\bibnamefont {Ryan-Anderson}}, \bibinfo {author} {\bibfnamefont {S.}~\bibnamefont {Sanders}}, \bibinfo {author} {\bibfnamefont {J.}~\bibnamefont {Sedlacek}}, \bibinfo {author}
  {\bibfnamefont {P.}~\bibnamefont {Shevchuk}}, \bibinfo {author} {\bibfnamefont {P.}~\bibnamefont {Siegfried}}, \bibinfo {author} {\bibfnamefont {T.}~\bibnamefont {Skripka}}, \bibinfo {author} {\bibfnamefont {B.}~\bibnamefont {Spaun}}, \bibinfo {author} {\bibfnamefont {R.}~\bibnamefont {Sprenkle}}, \bibinfo {author} {\bibfnamefont {R.}~\bibnamefont {Stutz}}, \bibinfo {author} {\bibfnamefont {M.}~\bibnamefont {Swallows}}, \bibinfo {author} {\bibfnamefont {R.}~\bibnamefont {Tobey}}, \bibinfo {author} {\bibfnamefont {A.}~\bibnamefont {Tran}}, \bibinfo {author} {\bibfnamefont {T.}~\bibnamefont {Tran}}, \bibinfo {author} {\bibfnamefont {E.}~\bibnamefont {Vogt}}, \bibinfo {author} {\bibfnamefont {C.}~\bibnamefont {Volin}}, \bibinfo {author} {\bibfnamefont {J.}~\bibnamefont {Walker}}, \bibinfo {author} {\bibfnamefont {A.}~\bibnamefont {Zolot}},\ and\ \bibinfo {author} {\bibfnamefont {J.}~\bibnamefont {Pino}},\ }\bibfield  {title} {\bibinfo {title} {A race-track trapped-ion quantum processor},\ }\href
  {https://doi.org/10.1103/physrevx.13.041052} {\bibfield  {journal} {\bibinfo  {journal} {Phys. Rev. X}\ }\textbf {\bibinfo {volume} {13}},\ \bibinfo {pages} {4} (\bibinfo {year} {2023})}\BibitemShut {NoStop}%
\bibitem [{\citenamefont {Jain}\ \emph {et~al.}(2024)\citenamefont {Jain}, \citenamefont {S\"{a}gesser}, \citenamefont {Hrmo}, \citenamefont {Torkzaban}, \citenamefont {Stadler}, \citenamefont {Oswald}, \citenamefont {Axline}, \citenamefont {Bautista-Salvador}, \citenamefont {Ospelkaus}, \citenamefont {Kienzler},\ and\ \citenamefont {Home}}]{Jain2024}%
  \BibitemOpen
  \bibfield  {author} {\bibinfo {author} {\bibfnamefont {S.}~\bibnamefont {Jain}}, \bibinfo {author} {\bibfnamefont {T.}~\bibnamefont {S\"{a}gesser}}, \bibinfo {author} {\bibfnamefont {P.}~\bibnamefont {Hrmo}}, \bibinfo {author} {\bibfnamefont {C.}~\bibnamefont {Torkzaban}}, \bibinfo {author} {\bibfnamefont {M.}~\bibnamefont {Stadler}}, \bibinfo {author} {\bibfnamefont {R.}~\bibnamefont {Oswald}}, \bibinfo {author} {\bibfnamefont {C.}~\bibnamefont {Axline}}, \bibinfo {author} {\bibfnamefont {A.}~\bibnamefont {Bautista-Salvador}}, \bibinfo {author} {\bibfnamefont {C.}~\bibnamefont {Ospelkaus}}, \bibinfo {author} {\bibfnamefont {D.}~\bibnamefont {Kienzler}},\ and\ \bibinfo {author} {\bibfnamefont {J.}~\bibnamefont {Home}},\ }\bibfield  {title} {\bibinfo {title} {Penning micro-trap for quantum computing},\ }\href {https://doi.org/10.1038/s41586-024-07111-x} {\bibfield  {journal} {\bibinfo  {journal} {Nature}\ }\textbf {\bibinfo {volume} {627}},\ \bibinfo {pages} {510–514} (\bibinfo {year} {2024})}\BibitemShut
  {NoStop}%
\bibitem [{\citenamefont {Shaw}\ \emph {et~al.}(2024{\natexlab{c}})\citenamefont {Shaw}, \citenamefont {Scholl}, \citenamefont {Finkelstein}, \citenamefont {Tsai}, \citenamefont {Choi},\ and\ \citenamefont {Endres}}]{shaw2023}%
  \BibitemOpen
  \bibfield  {author} {\bibinfo {author} {\bibfnamefont {A.~L.}\ \bibnamefont {Shaw}}, \bibinfo {author} {\bibfnamefont {P.}~\bibnamefont {Scholl}}, \bibinfo {author} {\bibfnamefont {R.}~\bibnamefont {Finkelstein}}, \bibinfo {author} {\bibfnamefont {R.~B.-S.}\ \bibnamefont {Tsai}}, \bibinfo {author} {\bibfnamefont {J.}~\bibnamefont {Choi}},\ and\ \bibinfo {author} {\bibfnamefont {M.}~\bibnamefont {Endres}},\ }\href@noop {} {\bibinfo {title} {Erasure-cooling, control, and hyper-entanglement of motion in optical tweezers}} (\bibinfo {year} {2024}{\natexlab{c}}),\ \Eprint {https://arxiv.org/abs/2311.15580} {arXiv:2311.15580 [quant-ph]} \BibitemShut {NoStop}%
\bibitem [{\citenamefont {Hou}\ \emph {et~al.}(2024)\citenamefont {Hou}, \citenamefont {Wu}, \citenamefont {Erickson}, \citenamefont {Cole}, \citenamefont {Zarantonello}, \citenamefont {Brandt}, \citenamefont {Geller}, \citenamefont {Kwiatkowski}, \citenamefont {Glancy}, \citenamefont {Knill}, \citenamefont {Wilson}, \citenamefont {Slichter},\ and\ \citenamefont {Leibfried}}]{Hou2024}%
  \BibitemOpen
  \bibfield  {author} {\bibinfo {author} {\bibfnamefont {P.-Y.}\ \bibnamefont {Hou}}, \bibinfo {author} {\bibfnamefont {J.~J.}\ \bibnamefont {Wu}}, \bibinfo {author} {\bibfnamefont {S.~D.}\ \bibnamefont {Erickson}}, \bibinfo {author} {\bibfnamefont {D.~C.}\ \bibnamefont {Cole}}, \bibinfo {author} {\bibfnamefont {G.}~\bibnamefont {Zarantonello}}, \bibinfo {author} {\bibfnamefont {A.~D.}\ \bibnamefont {Brandt}}, \bibinfo {author} {\bibfnamefont {S.}~\bibnamefont {Geller}}, \bibinfo {author} {\bibfnamefont {A.}~\bibnamefont {Kwiatkowski}}, \bibinfo {author} {\bibfnamefont {S.}~\bibnamefont {Glancy}}, \bibinfo {author} {\bibfnamefont {E.}~\bibnamefont {Knill}}, \bibinfo {author} {\bibfnamefont {A.~C.}\ \bibnamefont {Wilson}}, \bibinfo {author} {\bibfnamefont {D.~H.}\ \bibnamefont {Slichter}},\ and\ \bibinfo {author} {\bibfnamefont {D.}~\bibnamefont {Leibfried}},\ }\bibfield  {title} {\bibinfo {title} {Coherent coupling and non-destructive measurement of trapped-ion mechanical oscillators},\ }\href
  {https://doi.org/10.1038/s41567-024-02585-y} {\bibfield  {journal} {\bibinfo  {journal} {Nat. Phys.}\ } (\bibinfo {year} {2024})}\BibitemShut {NoStop}%
\bibitem [{\citenamefont {Royer}\ \emph {et~al.}(2022)\citenamefont {Royer}, \citenamefont {Singh},\ and\ \citenamefont {Girvin}}]{Royer2022}%
  \BibitemOpen
  \bibfield  {author} {\bibinfo {author} {\bibfnamefont {B.}~\bibnamefont {Royer}}, \bibinfo {author} {\bibfnamefont {S.}~\bibnamefont {Singh}},\ and\ \bibinfo {author} {\bibfnamefont {S.}~\bibnamefont {Girvin}},\ }\bibfield  {title} {\bibinfo {title} {Encoding qubits in multimode grid states},\ }\href {https://doi.org/10.1103/prxquantum.3.010335} {\bibfield  {journal} {\bibinfo  {journal} {PRX Quantum}\ }\textbf {\bibinfo {volume} {3}},\ \bibinfo {pages} {1} (\bibinfo {year} {2022})}\BibitemShut {NoStop}%
\bibitem [{\citenamefont {Liu}\ \emph {et~al.}(2024)\citenamefont {Liu}, \citenamefont {Singh}, \citenamefont {Smith}, \citenamefont {Crane}, \citenamefont {Martyn}, \citenamefont {Eickbusch}, \citenamefont {Schuckert}, \citenamefont {Li}, \citenamefont {Sinanan-Singh}, \citenamefont {Soley}, \citenamefont {Tsunoda}, \citenamefont {Chuang}, \citenamefont {Wiebe},\ and\ \citenamefont {Girvin}}]{Lui2024}%
  \BibitemOpen
  \bibfield  {author} {\bibinfo {author} {\bibfnamefont {Y.}~\bibnamefont {Liu}}, \bibinfo {author} {\bibfnamefont {S.}~\bibnamefont {Singh}}, \bibinfo {author} {\bibfnamefont {K.~C.}\ \bibnamefont {Smith}}, \bibinfo {author} {\bibfnamefont {E.}~\bibnamefont {Crane}}, \bibinfo {author} {\bibfnamefont {J.~M.}\ \bibnamefont {Martyn}}, \bibinfo {author} {\bibfnamefont {A.}~\bibnamefont {Eickbusch}}, \bibinfo {author} {\bibfnamefont {A.}~\bibnamefont {Schuckert}}, \bibinfo {author} {\bibfnamefont {R.~D.}\ \bibnamefont {Li}}, \bibinfo {author} {\bibfnamefont {J.}~\bibnamefont {Sinanan-Singh}}, \bibinfo {author} {\bibfnamefont {M.~B.}\ \bibnamefont {Soley}}, \bibinfo {author} {\bibfnamefont {T.}~\bibnamefont {Tsunoda}}, \bibinfo {author} {\bibfnamefont {I.~L.}\ \bibnamefont {Chuang}}, \bibinfo {author} {\bibfnamefont {N.}~\bibnamefont {Wiebe}},\ and\ \bibinfo {author} {\bibfnamefont {S.~M.}\ \bibnamefont {Girvin}},\ }\href@noop {} {\bibinfo {title} {Hybrid oscillator-qubit quantum processors: Instruction set
  architectures, abstract machine models, and applications}} (\bibinfo {year} {2024}),\ \Eprint {https://arxiv.org/abs/2407.10381} {arXiv:2407.10381} \BibitemShut {NoStop}%
\bibitem [{\citenamefont {Rymarz}\ \emph {et~al.}(2021)\citenamefont {Rymarz}, \citenamefont {Bosco}, \citenamefont {Ciani},\ and\ \citenamefont {DiVincenzo}}]{Rymarz2021}%
  \BibitemOpen
  \bibfield  {author} {\bibinfo {author} {\bibfnamefont {M.}~\bibnamefont {Rymarz}}, \bibinfo {author} {\bibfnamefont {S.}~\bibnamefont {Bosco}}, \bibinfo {author} {\bibfnamefont {A.}~\bibnamefont {Ciani}},\ and\ \bibinfo {author} {\bibfnamefont {D.~P.}\ \bibnamefont {DiVincenzo}},\ }\bibfield  {title} {\bibinfo {title} {Hardware-encoding grid states in a nonreciprocal superconducting circuit},\ }\href {https://doi.org/10.1103/physrevx.11.011032} {\bibfield  {journal} {\bibinfo  {journal} {Phys. Rev. X}\ }\textbf {\bibinfo {volume} {11}},\ \bibinfo {pages} {1} (\bibinfo {year} {2021})}\BibitemShut {NoStop}%
\bibitem [{\citenamefont {Duivenvoorden}\ \emph {et~al.}(2017)\citenamefont {Duivenvoorden}, \citenamefont {Terhal},\ and\ \citenamefont {Weigand}}]{Duivenvoorden2017}%
  \BibitemOpen
  \bibfield  {author} {\bibinfo {author} {\bibfnamefont {K.}~\bibnamefont {Duivenvoorden}}, \bibinfo {author} {\bibfnamefont {B.~M.}\ \bibnamefont {Terhal}},\ and\ \bibinfo {author} {\bibfnamefont {D.}~\bibnamefont {Weigand}},\ }\bibfield  {title} {\bibinfo {title} {Single-mode displacement sensor},\ }\href {https://doi.org/10.1103/physreva.95.012305} {\bibfield  {journal} {\bibinfo  {journal} {Phys. Rev. A}\ }\textbf {\bibinfo {volume} {95}},\ \bibinfo {pages} {1} (\bibinfo {year} {2017})}\BibitemShut {NoStop}%
\bibitem [{\citenamefont {Q-CTRL}(2023)}]{boulder_opal2}%
  \BibitemOpen
  \bibfield  {author} {\bibinfo {author} {\bibnamefont {Q-CTRL}},\ }\href@noop {} {\bibinfo {title} {Boulder {O}pal}},\ \bibinfo {howpublished} {https://q-ctrl.com/boulder-opal} (\bibinfo {year} {2023}),\ \bibinfo {note} {[Online]}\BibitemShut {NoStop}%
\bibitem [{\citenamefont {Nielsen}(2002)}]{Nielsen2002}%
  \BibitemOpen
  \bibfield  {author} {\bibinfo {author} {\bibfnamefont {M.~A.}\ \bibnamefont {Nielsen}},\ }\bibfield  {title} {\bibinfo {title} {A simple formula for the average gate fidelity of a quantum dynamical operation},\ }\href {https://doi.org/10.1016/s0375-9601(02)01272-0} {\bibfield  {journal} {\bibinfo  {journal} {Phys. Lett. A}\ }\textbf {\bibinfo {volume} {303}},\ \bibinfo {pages} {249–252} (\bibinfo {year} {2002})}\BibitemShut {NoStop}%
\bibitem [{\citenamefont {Carignan-Dugas}\ \emph {et~al.}(2019)\citenamefont {Carignan-Dugas}, \citenamefont {Wallman},\ and\ \citenamefont {Emerson}}]{CarignanDugas2019}%
  \BibitemOpen
  \bibfield  {author} {\bibinfo {author} {\bibfnamefont {A.}~\bibnamefont {Carignan-Dugas}}, \bibinfo {author} {\bibfnamefont {J.~J.}\ \bibnamefont {Wallman}},\ and\ \bibinfo {author} {\bibfnamefont {J.}~\bibnamefont {Emerson}},\ }\bibfield  {title} {\bibinfo {title} {Bounding the average gate fidelity of composite channels using the unitarity},\ }\href {https://doi.org/10.1088/1367-2630/ab1800} {\bibfield  {journal} {\bibinfo  {journal} {New J. Phys.}\ }\textbf {\bibinfo {volume} {21}},\ \bibinfo {pages} {053016} (\bibinfo {year} {2019})}\BibitemShut {NoStop}%
\bibitem [{\citenamefont {Pantaleoni}\ \emph {et~al.}(2020)\citenamefont {Pantaleoni}, \citenamefont {Baragiola},\ and\ \citenamefont {Menicucci}}]{Pantaleoni2020}%
  \BibitemOpen
  \bibfield  {author} {\bibinfo {author} {\bibfnamefont {G.}~\bibnamefont {Pantaleoni}}, \bibinfo {author} {\bibfnamefont {B.~Q.}\ \bibnamefont {Baragiola}},\ and\ \bibinfo {author} {\bibfnamefont {N.~C.}\ \bibnamefont {Menicucci}},\ }\bibfield  {title} {\bibinfo {title} {Modular bosonic subsystem codes},\ }\href {https://doi.org/10.1103/physrevlett.125.040501} {\bibfield  {journal} {\bibinfo  {journal} {Phys. Rev. Lett.}\ }\textbf {\bibinfo {volume} {125}},\ \bibinfo {pages} {4} (\bibinfo {year} {2020})}\BibitemShut {NoStop}%
\bibitem [{\citenamefont {White}\ \emph {et~al.}(2007)\citenamefont {White}, \citenamefont {Gilchrist}, \citenamefont {Pryde}, \citenamefont {O’Brien}, \citenamefont {Bremner},\ and\ \citenamefont {Langford}}]{White2007}%
  \BibitemOpen
  \bibfield  {author} {\bibinfo {author} {\bibfnamefont {A.~G.}\ \bibnamefont {White}}, \bibinfo {author} {\bibfnamefont {A.}~\bibnamefont {Gilchrist}}, \bibinfo {author} {\bibfnamefont {G.~J.}\ \bibnamefont {Pryde}}, \bibinfo {author} {\bibfnamefont {J.~L.}\ \bibnamefont {O’Brien}}, \bibinfo {author} {\bibfnamefont {M.~J.}\ \bibnamefont {Bremner}},\ and\ \bibinfo {author} {\bibfnamefont {N.~K.}\ \bibnamefont {Langford}},\ }\bibfield  {title} {\bibinfo {title} {Measuring two-qubit gates},\ }\href {https://doi.org/10.1364/josab.24.000172} {\bibfield  {journal} {\bibinfo  {journal} {J. Opt. Soc. Am. B}\ }\textbf {\bibinfo {volume} {24}},\ \bibinfo {pages} {172} (\bibinfo {year} {2007})}\BibitemShut {NoStop}%
\bibitem [{\citenamefont {Bialczak}\ \emph {et~al.}(2010)\citenamefont {Bialczak}, \citenamefont {Ansmann}, \citenamefont {Hofheinz}, \citenamefont {Lucero}, \citenamefont {Neeley}, \citenamefont {O’Connell}, \citenamefont {Sank}, \citenamefont {Wang}, \citenamefont {Wenner}, \citenamefont {Steffen}, \citenamefont {Cleland},\ and\ \citenamefont {Martinis}}]{Bialczak2010}%
  \BibitemOpen
  \bibfield  {author} {\bibinfo {author} {\bibfnamefont {R.~C.}\ \bibnamefont {Bialczak}}, \bibinfo {author} {\bibfnamefont {M.}~\bibnamefont {Ansmann}}, \bibinfo {author} {\bibfnamefont {M.}~\bibnamefont {Hofheinz}}, \bibinfo {author} {\bibfnamefont {E.}~\bibnamefont {Lucero}}, \bibinfo {author} {\bibfnamefont {M.}~\bibnamefont {Neeley}}, \bibinfo {author} {\bibfnamefont {A.~D.}\ \bibnamefont {O’Connell}}, \bibinfo {author} {\bibfnamefont {D.}~\bibnamefont {Sank}}, \bibinfo {author} {\bibfnamefont {H.}~\bibnamefont {Wang}}, \bibinfo {author} {\bibfnamefont {J.}~\bibnamefont {Wenner}}, \bibinfo {author} {\bibfnamefont {M.}~\bibnamefont {Steffen}}, \bibinfo {author} {\bibfnamefont {A.~N.}\ \bibnamefont {Cleland}},\ and\ \bibinfo {author} {\bibfnamefont {J.~M.}\ \bibnamefont {Martinis}},\ }\bibfield  {title} {\bibinfo {title} {Quantum process tomography of a universal entangling gate implemented with josephson phase qubits},\ }\href {https://doi.org/10.1038/nphys1639} {\bibfield  {journal} {\bibinfo
  {journal} {Nat. Phys.}\ }\textbf {\bibinfo {volume} {6}},\ \bibinfo {pages} {409–413} (\bibinfo {year} {2010})}\BibitemShut {NoStop}%
\bibitem [{\citenamefont {Riofrío}\ \emph {et~al.}(2017)\citenamefont {Riofrío}, \citenamefont {Gross}, \citenamefont {Flammia}, \citenamefont {Monz}, \citenamefont {Nigg}, \citenamefont {Blatt},\ and\ \citenamefont {Eisert}}]{Riofro2017}%
  \BibitemOpen
  \bibfield  {author} {\bibinfo {author} {\bibfnamefont {C.~A.}\ \bibnamefont {Riofrío}}, \bibinfo {author} {\bibfnamefont {D.}~\bibnamefont {Gross}}, \bibinfo {author} {\bibfnamefont {S.~T.}\ \bibnamefont {Flammia}}, \bibinfo {author} {\bibfnamefont {T.}~\bibnamefont {Monz}}, \bibinfo {author} {\bibfnamefont {D.}~\bibnamefont {Nigg}}, \bibinfo {author} {\bibfnamefont {R.}~\bibnamefont {Blatt}},\ and\ \bibinfo {author} {\bibfnamefont {J.}~\bibnamefont {Eisert}},\ }\bibfield  {title} {\bibinfo {title} {Experimental quantum compressed sensing for a seven-qubit system},\ }\href {https://doi.org/10.1038/ncomms15305} {\bibfield  {journal} {\bibinfo  {journal} {Nat. Comms.}\ }\textbf {\bibinfo {volume} {8}},\ \bibinfo {pages} {1} (\bibinfo {year} {2017})}\BibitemShut {NoStop}%
\bibitem [{\citenamefont {Chenu}\ \emph {et~al.}(2017)\citenamefont {Chenu}, \citenamefont {Beau}, \citenamefont {Cao},\ and\ \citenamefont {del Campo}}]{Chenu2017}%
  \BibitemOpen
  \bibfield  {author} {\bibinfo {author} {\bibfnamefont {A.}~\bibnamefont {Chenu}}, \bibinfo {author} {\bibfnamefont {M.}~\bibnamefont {Beau}}, \bibinfo {author} {\bibfnamefont {J.}~\bibnamefont {Cao}},\ and\ \bibinfo {author} {\bibfnamefont {A.}~\bibnamefont {del Campo}},\ }\bibfield  {title} {\bibinfo {title} {Quantum simulation of generic many-body open system dynamics using classical noise},\ }\href {https://doi.org/10.1103/physrevlett.118.140403} {\bibfield  {journal} {\bibinfo  {journal} {Phys. Rev. Lett.}\ }\textbf {\bibinfo {volume} {118}},\ \bibinfo {pages} {14} (\bibinfo {year} {2017})}\BibitemShut {NoStop}%
\bibitem [{\citenamefont {Gu}\ and\ \citenamefont {Franco}(2019)}]{Gu2019}%
  \BibitemOpen
  \bibfield  {author} {\bibinfo {author} {\bibfnamefont {B.}~\bibnamefont {Gu}}\ and\ \bibinfo {author} {\bibfnamefont {I.}~\bibnamefont {Franco}},\ }\bibfield  {title} {\bibinfo {title} {When can quantum decoherence be mimicked by classical noise?},\ }\href {https://doi.org/10.1063/1.5099499} {\bibfield  {journal} {\bibinfo  {journal} {J. Chem. Phys.}\ }\textbf {\bibinfo {volume} {151}},\ \bibinfo {pages} {1} (\bibinfo {year} {2019})}\BibitemShut {NoStop}%
\end{thebibliography}%

\section*{Supplementary Information}

\setcounter{figure}{0}
\renewcommand{\figurename}{Supplemental figure}
\renewcommand{\thefigure}{SI\arabic{figure}}
\renewcommand{\theHfigure}{SI\arabic{figure}}

\setcounter{equation}{0}
\renewcommand{\theequation}{SI\arabic{equation}}

\subsection{GKP States} \label{sup:GKP-states}
Finite-energy GKP states approximate the ideal code words, which are represented by an equal superposition of coherent states: $\ket{\mu_\mathrm{L,id.}} = \sum_{k,l = -\infty}^\infty e^{-i\pi( kl+\mu l/2)}\ket{(2k+\mu)\alpha + l\beta}$, where $\mu=0$ ($\mu=1$) parameterise the logical $\ket{+Z_\mathrm{L}}(\ket{-Z_\mathrm{L}})$ state, and $\beta =i\alpha =i\sqrt{\pi/2}$ define the square-lattice spacing. These code words are stabilised by the commuting operators $\hat{S}_X = \hat{D}(2\alpha)$ and $\hat{S}_Z = \hat{D}(2\beta)$, and admit logical Pauli operators $\hat{X}_\mathrm{L} = \hat{D}(\alpha)$, $\hat{Y}_\mathrm{L} = \hat{D}(\alpha+\beta)$ and $\hat{Z}_\mathrm{L} = \hat{D}(\beta)$. Here, $\hat{D}(\gamma) = e^{\gamma \hat{a}^\dagger - \gamma^* \hat{a}}$ denotes the bosonic displacement operator. Finite-energy GKP states incorporate a Gaussian envelope to the ideal GKP states such that $\ket{\mu_\mathrm{L}} = e^{-\Delta^2 \hat{a}^\dagger \hat{a}}\ket{\mu_\mathrm{L,id.}}$, where $\Delta$ parameterises the Gaussian envelope. The finite-energy GKP states, $\ket{\mu_\mathrm{L}}$, are only approximately stabilised by $\hat{S}_X$ and $\hat{S}_Z$, approaching the ideal behaviour as $\Delta \rightarrow 0$.

In addition to the above definition of the GKP code states, one may define the code states to be the quasi-degenerate ground states of the Hamiltonian~\cite{Rymarz2021,Kolesnikow2024}
\begin{equation}
    \hat{H}_{\mathrm{GKP}} = \omega_0\hat{a}^\dagger \hat{a}-J(\cos (2 \sqrt{\pi} \hat{x})+\cos (2 \sqrt{\pi} \hat{p})), \label{eqn:GKP-ham}
\end{equation}
where $\hat{x} = (\hat{a} + \hat{a}^\dagger)/\sqrt{2}$ and $\hat{p} = -i(\hat{a} - \hat{a}^\dagger)/\sqrt{2}$ are the position and momentum operators for a single bosonic mode, and $\omega_0$ and $J$ define the energy scale. In the limit of large $J/\omega_0$, the ground states are approximately logical Hadamard eigenstates in the $\ket{\pm Z_\mathrm{L}}$ basis~\cite{Rymarz2021},
\begin{subequations}
\begin{align}
    \ket{+H_\mathrm{L}} &= \cos{(\pi/8)} \ket{+Z_\mathrm{L}} + \sin{(\pi/8)}\ket{-Z_\mathrm{L}},\\
    \ket{-H_\mathrm{L}} &= -\sin{(\pi/8)} \ket{+Z_\mathrm{L}} + \cos{(\pi/8)}\ket{-Z_\mathrm{L}},
\end{align}
\label{eqn:Hpm}
\end{subequations}
and the squeezing parameter $\Delta$ of these states is related to the energy scales through $\Delta = (\omega_0/(4\pi J))^{1/4}$. We find numerical diagonalisation of Eq.~\ref{eqn:GKP-ham} to be an efficient procedure to obtain the finite-energy GKP code words numerically. The relations in Eq.~\ref{eqn:Hpm} may be inverted to obtain the $\ket{\pm Z_\mathrm{L}}$ code words, and the other logical Pauli eigenstates obtained from the appropriate superpositions of these. 

Furthermore, the squeezing parameter of these states in each quadrature $\Delta_{X/Z}$ may be independently calculated from their stabiliser expectation values through the relations~\cite{Duivenvoorden2017}
\begin{subequations}
\begin{align}
    \Delta_{X} &= \sqrt{ - \dfrac{1}{2\pi} \log{\left(  | \langle \hat{S}_{X} \rangle |^2\right)}},\\
    \Delta_{Z} &= \sqrt{ - \dfrac{1}{2\pi} \log{\left(  | \langle \hat{S}_{Z} \rangle |^2\right)}},
\end{align}
\end{subequations}
which can also be expressed in units of dB with $\Delta_{X/Z}~(\mathrm{dB}) = -10 \log_{10}(\Delta_{X/Z}^2)$. In this work, we set $J/\omega_0=5.95$ in Eq.~\ref{eqn:GKP-ham} to target logical states $\{\ket{+Z_\mathrm{L}}, \ket{-Z_\mathrm{L}}, \ket{+X_\mathrm{L}}, \ket{+Y_\mathrm{L}}\}$ with squeezing parameters $[\Delta_X,\Delta_Z]$ of $\{[8.39,7.90],[8.36,8.88],[7.90,8.39],[8.38,8.38]\}~\mathrm{dB}$.

\subsection{Numerical Optimisation}\label{sup:num_op}

The GKP state preparation, single-qubit gates, two-qubit $\mathrm{CZ}_\mathrm{L}$ gate and Bell state preparation pulses are numerically optimised using Q-CTRL's graph-based optimiser (Boulder Opal \cite{boulder_opal1, boulder_opal2}). We choose a model-based optimisation approach in which the target Hamiltonian is represented as a computational graph. Each output pulse is depicted by a complex-valued piecewise-constant function, with each segment serving as an optimisable variable. We optimise the parameters of a control Hamiltonian which describes a laser-induced spin-boson interaction on a single trapped ion. The Hamiltonian is composed of two terms referred to as the red- and blue-sideband interactions. To minimise errors from otherwise neglected terms in the Lamb-Dicke regime~\cite{Wineland1998}, we consider up to third-order terms of the Lamb-Dicke expansion and remove fast oscillating terms by making a rotating wave approximation,
\begin{align}
    \hat{H}_j(t) & = \frac{\Omega_j}{2} \hat{\sigma}^{+}  \left(\hat{a}_j -\frac{\eta_j^2}{2} \hat{a}_j\hat{a}_j^\dagger \hat{a}_j \right)e^{i \phi_\mathrm{r}(t)} \nonumber\\
    &+ \frac{\Omega_j}{2} \hat{\sigma}^{+}  \left(\hat{a}^{\dagger}_j - \frac{\eta_j^2}{2} \hat{a}^\dagger_j \hat{a}_j \hat{a}^\dagger_j \right)e^{i \phi_\mathrm{b}(t)}  +\text{h.c.},
    \label{eqn:ModifiedControlHamiltonian}
\end{align}
where $\hat{\sigma}^+ = \ket{\uparrow}\bra{\downarrow}$ and $\Omega_j$ is the Rabi rate of the first-order red- and blue-sideband interactions. $\phi_\mathrm{r}$ and $\phi_\mathrm{b}$ are red- and blue-sideband phases that are related to the spin and motional phases of Eq.~\ref{Eq:ControlHamiltonian} via $\phi_\mathrm{s} = (\phi_\mathrm{r}+\phi_\mathrm{b})/2$, $\phi_\mathrm{m} = (\phi_\mathrm{r} -\phi_\mathrm{b})/2$. $\eta$ is the Lamb-Dicke parameter, and takes the values $\eta_x =0.083$ and $\eta_y =0.078$ for the $x$- and $y$- motional modes, respectively. We set $\eta=\eta_x$ for both modes to reduce the number of compiled waveforms. The Hilbert space of each bosonic mode is truncated at $N_\mathrm{Fock} = 50$ Fock states, which leads to negligible infidelities for the moderately squeezed states considered here. The control Hamiltonian in Eq. \ref{Eq:ControlHamiltonian} is recovered only when considering terms that are linear in $\hat{a}$ and $\hat{a}^\dagger$ from the above Hamiltonian.

The phases $\phi_\mathrm{r}$ and $\phi_\mathrm{b}$ are represented as piecewise constant functions by the numerical optimiser, and are parameterised by $N_\mathrm{opt}$ optimisable segments. We impose additional constraints on the pulses, which are modeled as nodes in the optimisation graph, to simplify their experimental implementation and to avoid signal distortions from the finite bandwidth of the acousto-optics modulator (AOM) used to modulate the phases. First, we limit the slew rates of the $N_\mathrm{opt}$ phases, which limits the maximum rate of change between two adjacent segments. The slew rate, $\mathrm{SR} =\max \left|d \phi_{\mathrm{r,b}}(t) / d t \right|$, is heuristically determined. Second, we apply a sinc filter to the $N_\mathrm{opt}$ optimisable segments, and resample $N_\mathrm{seg}$ filtered segments that make up the final waveform. We set the sinc filter's cutoff frequency to $f_c  = 2\pi \times 35 \operatorname{KHz}$, chosen to be much smaller than the AOM's bandwidth. Third, we set $\phi_\mathrm{r}$ and $\phi_\mathrm{b}$ to be zero at $t=0$ to avoid phase discontinuities at the beginning of the pulse.

The cost function, $\mathcal{C}$, is encoded as a graph node and is defined as
\begin{equation}
    \label{eq:cost_function}
    \mathcal{C} =  (1-\mathcal{F}) + \epsilon\dfrac{T}{T_\mathrm{max}},
\end{equation}
where $\mathcal{F}$ is the fidelity and is uniquely defined for each optimisation problem. The convergence criterion, $\epsilon$, is chosen such that the optimiser minimises the duration of the pulse while ensuring that $1-\mathcal{F} \lesssim \epsilon$. $T$ is the duration of the optimised pulse and $T_\mathrm{max}$ is the maximum allowable pulse duration in the optimisation. 

\subsubsection{GKP state preparation graph optimisation}

We find four numerically optimised pulses to prepare each of the logical GKP states, $\ket{\psi_\mathrm{L}} \in \{\ket{+Z_\mathrm{L}}, \ket{-Z_\mathrm{L}},\ket{+X_\mathrm{L}},\ket{+Y_\mathrm{L}}\}$.
We target a unitary evolution, $\hat{U}_\mathrm{prep}$, that prepares the ideal qubit-bosonic state from vacuum, $\ket{\downarrow, \psi_{\mathrm{L}}} = \hat{U}_\mathrm{prep} \ket{\downarrow, 0}$. The fidelity used in the cost function of Eq.~\ref{eq:cost_function} is defined as the overlap of the evolved state with the target state
\begin{equation}
    \label{eq:fidelity_prep}
    \mathcal{F} = |\bra{\downarrow, \psi_{\mathrm{L}}} \hat{U} \ket{\downarrow, 0}|^2.
\end{equation}
The maximum allowed duration is set to $T_\mathrm{max} = \SI{2}{ms}$ and the convergence criterion $\epsilon = 0.05$. We set $N_\mathrm{opt} = 90$ optimisable segments with a slew rate $\mathrm{SR} \times T = 2\pi \times 60 $ and resample $N_\mathrm{seg} = 240$ segments after filtering. The numerically optimised waveforms to prepare the states $\{\ket{+Z_\mathrm{L}}, \ket{-Z_\mathrm{L}},\ket{+X_\mathrm{L}},\ket{+Y_\mathrm{L}}\}$ are shown in Fig \ref{fig:prep_pulses}a-d. The state fidelities of all waveforms calculated from Eq.~\ref{eq:fidelity_prep} are $\mathcal{F} > 0.995$.

The optimised waveform for the $\ket{+Z_\mathrm{L}}$ state can, in principle, be reused to prepare the $\ket{+X_\mathrm{L}}$ state after the transformations $\phi_\mathrm{b} \rightarrow \phi_\mathrm{b} + \pi/2$ and $\phi_\mathrm{r} \rightarrow \phi_\mathrm{r} - \pi/2$, since $\ket{+Z_\mathrm{L}}$ and $\ket{+X_\mathrm{L}}$ are related by a $\pi/2$ phase space rotation. However, we choose to separately optimise a waveform to prepare $\ket{+X_\mathrm{L}}$ to ensure that $\phi_\mathrm{r} = \phi_\mathrm{b} = 0$ at $t=0$.

\subsubsection{Single-qubit GKP gate graph optimisation}

We obtain three numerically optimised pulses to implement each single-qubit gate, $\hat{U}_\mathrm{SQ} \in \{\hat{R}_\mathrm{L}^X(-\pi/2), \hat{R}_\mathrm{L}^Z(-\pi/2), \hat{T}_\mathrm{L}\}$. The operator $\hat{U}_\mathrm{SQ}$, which acts on the joint spin-boson Hilbert space, maps a logical input state, $\ket{i_\mathrm{L}} \in \{\ket{\pm X_\mathrm{L}}, \ket{\pm Y_\mathrm{L}}, \ket{\pm Z_\mathrm{L}}\}$, to a targeted logical output state, $\ket{t_\mathrm{L}}$, such that $\ket{\downarrow, t_\mathrm{L}} = \hat{U}_\mathrm{SQ}\ket{\downarrow, i_\mathrm{L}}$. The target states $\ket{t_{\mathrm{L}}}$ are found from the action of an ideal gate acting on a two-level system. For example, for an initial state $\ket{i_\mathrm{L}} = \ket{+Z_\mathrm{L}}$, the $\hat{R}^X_\mathrm{L}(-\pi/2)$ gate gives the target state $\ket{t_\mathrm{L}} = \ket{+Y_\mathrm{L}}$.
This definition ensures that the operation preserves the finite-energy envelope of the approximate GKP states. The fidelity in Eq.~\ref{eq:cost_function} is then defined as the state fidelity averaged over the six initial Pauli eigenstates,
\begin{equation}
    \label{eq:fidelity_1q}
    \mathcal{F} = \frac{1}{6}\sum_{m}|\bra{\downarrow, t_{\mathrm{L}, m}} \hat{U} \ket{\downarrow, i_{\mathrm{L}, m}}|^2,
\end{equation}
where the index $m$ enumerates the six Pauli eigenstates. We set the maximum allowed duration to $T_\mathrm{max} =\SI{600}{\micro s}$ and the convergence criterion to $\epsilon=0.01$. We optimise $\phi_\mathrm{r}, \phi_\mathrm{b}$ for $N_\mathrm{opt} = 90$ optimisable segments with a slew rate $\mathrm{SR} \times T = 2\pi \times 60$ and resample $N_\mathrm{seg} = 240$ segments. The numerically optimised waveforms to implement the single-qubit gates are shown in Fig.~\ref{fig:gate_pulses}a-c.

\subsubsection{Two-qubit GKP gate graph optimisation}

\begin{figure*}
    \centering
   \includegraphics[]{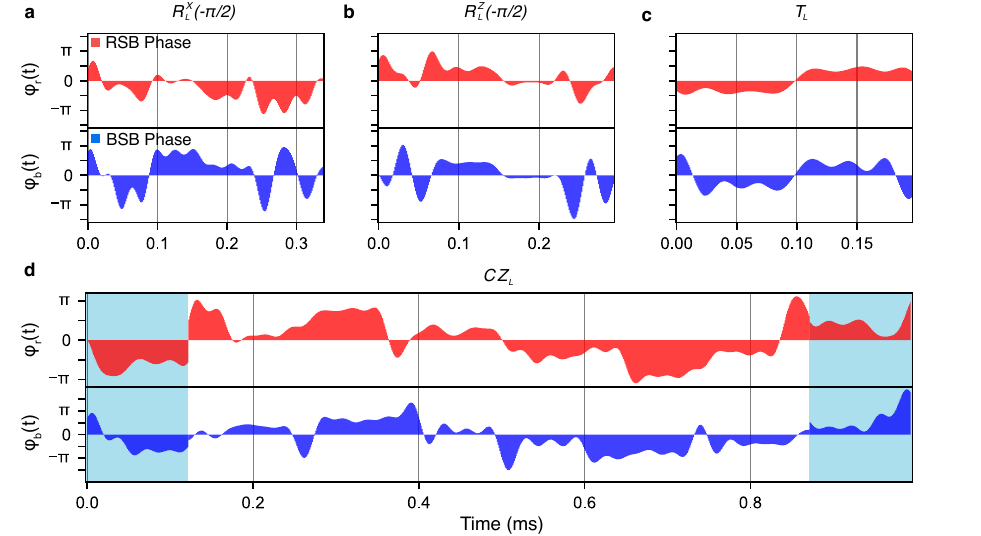}
    \caption{\textbf{Gate pulses.} 
    Numerically optimised red-sideband phases $\phi_\mathrm{r}(t)$ (red) and blue-sideband phases $\phi_\mathrm{b}(t)$ (blue) for the gate pulses featured in this Letter. Using the Hamiltonian from Eq. \ref{eqn:ModifiedControlHamiltonian} and $\Omega_{\mathrm{r},\mathrm{b}} = 2\pi \times \SI{2.4}{kHz}$, each waveform is used to produce the target logical gates denoted by: \textbf{a}, $\hat{R}_\mathrm{L}^X(-\pi/2)$,
    \textbf{b}, $\hat{R}_\mathrm{L}^Z(-\pi/2)$,
    \textbf{c}, $\hat{T}_\mathrm{L}$,
    \textbf{d}, $\mathrm{CZ}_\mathrm{L}$.  For the $\mathrm{CZ}_\mathrm{L}$ pulse, the blue shaded (unshaded) regions indicate the part of the SDF interaction applied to the radial-$y$ mode (radial-$x$ mode).}
    \label{fig:gate_pulses}
\end{figure*}

The logical two-qubit $\mathrm{CZ}_\mathrm{L}$ gate is obtained by numerically optimising two unitaries, $\hat{U}_{1}$ and $\hat{U}_{2}$. $\hat{U}_1$ ($\hat{U}_2$) acts on the first (second) bosonic mode and its interaction is obtained by setting $j_1=y$ ($j_2=x$) in Eq.~\ref{eqn:ModifiedControlHamiltonian}. The $\mathrm{CZ}_\mathrm{L}$ operator, $\hat{U}_\mathrm{CZ} = \hat{U}_1^\dagger \hat{U}_2 \hat{U}_1$, is then defined as the mapping $\left\{\left| \downarrow,i_{L},k_{L}\right\rangle\right\} \rightarrow\left\{\left|\downarrow,t_{L}\right\rangle\right\}$, where $\ket{i_\mathrm{L}}, \ket{k_\mathrm{L}} \in \{\ket{\pm X_\mathrm{L}}, \ket{\pm Y_\mathrm{L}}, \ket{\pm Z_\mathrm{L}}\}$. The target states $\ket{t_\mathrm{L}}$ are found from the action of an ideal CZ operation on a joint two-level system. For example, the initial state $\ket{i_\mathrm{L}, k_\mathrm{L}} = \ket{-Z_\mathrm{L}, +X_\mathrm{L}}$ gives the target state $\ket{t_\mathrm{L}} = \ket{-Z_\mathrm{L}, -X_\mathrm{L}}$. The fidelity of the cost function, $\mathcal{F}$, is defined as the state fidelity averaged over the $36$ possible initial two-qubit Pauli eigenstates,
\begin{equation}
    \label{eq:fidelity_2q}
    \mathcal{F} = \frac{1}{36}\sum_{m,n}|\bra{\downarrow,t_{L,m, n} } \hat{U}_{1}^\dagger \hat{U}_{2} \hat{U}_{1} \ket{\downarrow,  i_{L, m},k_{L, n}}|^2,
\end{equation}
where the indicies $m,n$ label the different Pauli eigenstates. The unitaries $\hat{U}_1$ and $\hat{U}_2$ are calculated from $\hat{U}_{1} = \hat{\mathcal{T}} e^{-i \int^{T_1}_0 \hat{H}_1(t) dt}$ and $\hat{U}_{2} = \hat{\mathcal{T}} e^{-i \int^{T_2}_0 \hat{H}_2(t) dt}$, respectively, where $\hat{\mathcal{T}}$ is the time-ordering operator. This results in a total duration $T = 2T_1 + T_2$. The maximum allowable duration is set to $T_\mathrm{max} = 2 \textrm{ ms}$ and the convergence criterion to $\epsilon = 0.05$. The operator $\hat{U}_1$ ($\hat{U}_2$) is obtained from $N_{\mathrm{opt},1} = 30$ ($N_{\mathrm{opt},2} = 270$) optimisable segments and $N_{\mathrm{seg},1} = 120$ ($N_{\mathrm{seg},2} = 720$) segments are resampled after filtering. The slew rates for $\hat{U}_1$ and $\hat{U}_2$ are $\mathrm{SR}_1 \times T = 2 \pi \times 20$ and $\mathrm{SR}_2 \times T = 2 \pi \times 80$, respectively. The numerically optimised waveforms to implement the $\mathrm{CZ}_\mathrm{L}$ gate is shown in Fig.~\ref{fig:gate_pulses}d.

\subsubsection{GKP Bell state preparation graph optimisation}

\begin{figure*}
    \centering
   \includegraphics[]{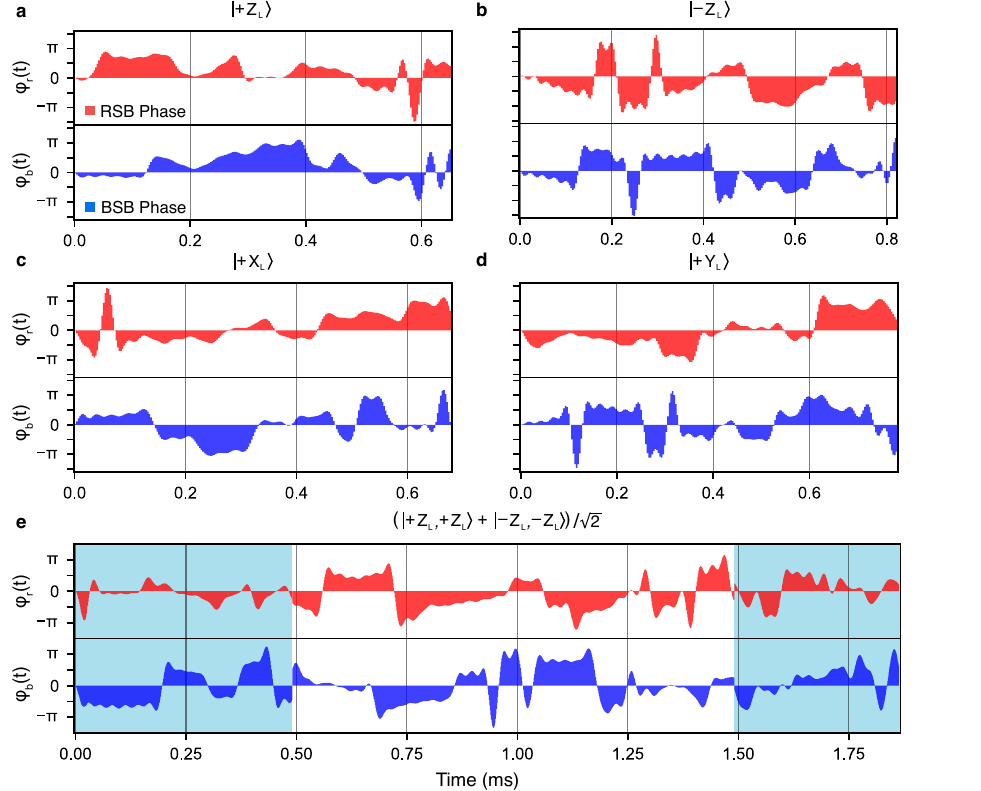}
    \caption{\textbf{State preparation pulses.} 
    Numerically optimised red-sideband phases $\phi_\mathrm{r}(t)$ (red) and blue-sideband phases $\phi_\mathrm{b}(t)$ (blue) for the state preparation pulses featured in this Letter. Using the Hamiltonian from Eq. \ref{eqn:ModifiedControlHamiltonian} and $\Omega_{\mathrm{r,b}} = 2\pi \times \SI{2.4}{kHz}$, each waveform is used to produce the target logical states denoted by: \textbf{a}, $\ket{+Z_\mathrm{L}}$,
    \textbf{b}, $\ket{-Z_\mathrm{L}}$,
    \textbf{c}, $\ket{+X_\mathrm{L}}$,
    \textbf{d}, $\ket{+Y_\mathrm{L}}$, and
    \textbf{e}, Logical Bell state $\ket{\Phi^+_\mathrm{L}} = \left(\ket{+Z_\mathrm{L}, +Z_\mathrm{L}} +\ket{-Z_\mathrm{L}, -Z_\mathrm{L}}\right)/\sqrt{2}$. For the Bell state pulse, the blue shaded (unshaded) regions indicate the part of the SDF interaction applied to the radial-$y$ mode (radial-$x$ mode).
    }
    \label{fig:prep_pulses}
\end{figure*}

A numerically optimised pulse is obtained to prepare the Bell state, $\ket{\psi_\mathrm{target}} = \left(\ket{+Z_\mathrm{L} ,+Z_\mathrm{L}} +\ket{-Z_\mathrm{L}, -Z_\mathrm{L}}\right)/\sqrt{2}$, from vacuum. We target the operator, $\hat{U}_\mathrm{Bell}$, such that $\ket{\downarrow, \psi_\mathrm{target}} = \hat{U}_\mathrm{Bell} \ket{\downarrow, 0, 0}$. We decompose $\hat{U}_\mathrm{Bell}$ into three unitary operators, $\hat{U}_\mathrm{Bell} = \hat{U}_3 \hat{U}_2 \hat{U}_1$, such that $\hat{U}_{1,3}$ act on mode $y$ and $\hat{U}_2$ on mode $x$. We define a state fidelity,
\begin{equation}
    \label{eq:fidelity_bell}
    \mathcal{F} = |\bra{\downarrow,\psi_\mathrm{target}}\hat{U}_{3} \hat{U}_{2} \hat{U}_{1} \ket{\downarrow, 0,0}|^2,
\end{equation}
where the unitary evolution of the state preparation pulse is given by $\hat{U}_{m} = \hat{\mathcal{T}}e^{-i \int^{T_m}_0 \hat{H}_{j_m}(t) dt}$, where $m \in \{1,2,3\}$ labels the unitary and $j_m$ labels the mode, with $j_1 = j_3 = y$ and $j_2 = x$. The total duration is given by $T = \sum_m T_m$, and we set the maximum allowable duration to $T_\mathrm{max} = 2 \textrm{ ms}$ and the convergence criterion to $\epsilon = 0.05$. The slew rates for $\{\hat{U}_1, \hat{U}_2, \hat{U}_3\}$ are set to $\mathrm{SR} \times T = 2 \pi \times \{30, 60, 30\}$. We also set $N_\mathrm{opt}=\{45, 90, 45\}$ optimisation segments, and $N_\mathrm{seg}=\{400, 800, 400\}$ resampled segments. The numerically optimised waveform to prepare the Bell state is shown in Fig \ref{fig:prep_pulses}e. The state fidelity calculated from Eq.~\ref{eq:fidelity_bell} of the Bell state prepared from the numerically optimised pulse is $\mathcal{F} = 0.984$.

\subsubsection{Gate characterisation}

We characterise the quality of the gate operation, $\hat{U}_\mathrm{gate}$, obtained from the single-qubit and two-qubit optimised gate pulses by computing their \textit{average gate fidelity}~\cite{Eickbusch2022,Nielsen2002,CarignanDugas2019},
\begin{equation}
    \mathcal{F}_\mathrm{gate} = \dfrac{1}{d(d+1)}\operatorname{Tr} \left(R^T[\hat{U}_\mathrm{ideal}] R_\mathrm{L} [\hat{U}_\mathrm{gate}] \right) + \dfrac{1}{d+1},
    \label{eq:av_gate_fid}
\end{equation}
where $d=2~(d=4)$ for a single-qubit (two-qubit) gate. Here, $R$ is the Pauli Transfer Matrix (PTM) with elements $R_{ij} = \frac{1}{d} \operatorname{Tr}\left(\hat{E}_i \hat{U}_\mathrm{ideal} \hat{E}_j \hat{U}_\mathrm{ideal}^\dagger\right)$, where $\hat{U}_\mathrm{ideal}$ acts on a two-level system. For single-qubit gates, $\hat{E}_i  \in \{\hat{I}, \hat{\sigma}_x, \hat{\sigma}_y,\hat{\sigma}_z\}$ and, for two-qubit gates, $\hat{E}_i  \in \{\hat{I}, \hat{\sigma}_x, \hat{\sigma}_y,\hat{\sigma}_z\}^{\otimes 2}$, where $\hat{\sigma}_i$ are the usual single-qubit Pauli operators. The ideal operators are $\hat{U}_\mathrm{ideal} \in \{\hat{R}^X(-\pi/2), \hat{R}^Z(-\pi/2), \hat{T}\}$ for the single-qubit gates and $\hat{U}_\mathrm{ideal} = \mathrm{CZ}$ for the two-qubit gate. The matrix elements $R_{\mathrm{L},ij} =\frac{1}{d} \operatorname{Tr}\left(\hat{E}_{\Delta,i} \hat{U}_\mathrm{gate} \hat{E}_{\Delta,j} \hat{U}_\mathrm{gate}^\dagger\right)$ of the logical PTM are similarly defined, however $\hat{U}_\mathrm{gate}$ and $\hat{E}_{\Delta,i}$ span the spin and bosonic mode(s). The finite-energy Pauli operators $\hat{E}_{\Delta,i}$ are defined with respect to finite energy GKP states obtained from section~\ref{sup:GKP-states}. For example, we can define logical two-qubit Pauli operators as $\hat{X}_\Delta \otimes \hat{X}_\Delta = \ket{\downarrow} \bra{\downarrow} \otimes \left( \ket{+Z_\mathrm{L}}\bra{-Z_\mathrm{L}} + \ket{-Z_\mathrm{L}}\bra{+Z_\mathrm{L}} \right)^{\otimes 2}$, with similar expressions using other logical finite-energy operators $\hat{I}_\Delta, \hat{Y}_\Delta, \hat{Z}_\Delta$. 

The numerically optimised single-qubit gates result in average gate fidelities calculated from Eq.~\ref{eq:av_gate_fid} of $\mathcal{F}_\mathrm{gate} > 0.998$. The numerically optimised two-qubit gate has an average gate fidelity of $\mathcal{F}_\mathrm{gate} =0.990$.

\subsection{Logical Measurements On The GKP Code Using The Stabiliser Sub-System Decomposition}\label{sup:log_meas}
The stabiliser sub-system decomposition (SSSD) provides a way to divide the infinite-dimensional Hilbert space into a logical subsystem, containing the logical information encoding in a GKP state, and a gauge subsystem, referred to as the stabiliser subsystem~\cite{Shaw2024,Pantaleoni2020}. Here, we make use of this formalism to accurately and efficiently readout the logical information encoded in our states.

Logical readout for the ideal square GKP code corresponds to measuring the expectation values of the logical Pauli operators,
\begin{equation}
    \hat{X}_\mathrm{L}  = \hat{D}(\ell_\mathrm{s}),~
    \hat{Y}_\mathrm{L}  = \hat{D}(\ell_\mathrm{s}+i\ell_\mathrm{s}),~
    \hat{Z}_\mathrm{L} = \hat{D}(i \ell_\mathrm{s}),
    \label{eq:pauli_logical_op}
\end{equation}
where $\ell_\mathrm{s} = \sqrt{2\pi}$ is the lattice spacing for the square GKP code. The expectation values of these logical Pauli operators are obtained by applying an SDF interaction, which transfers information from the bosonic mode to the ancilla, followed by an ancilla measurement. Setting the phases $\phi_\mathrm{s}$ and $\phi_\mathrm{m}$ to be time-independent with $\phi_\text{s} = 0$, the application of the SDF interaction from Eq. \ref{Eq:ControlHamiltonian} for a time $t$ realises the operation, $\hat{D}(\gamma \hat{\sigma}_x / 2)$, where $\gamma=-i\Omega t e^{-i \phi_\mathrm{m} }$. The magnitude of the displacement, $|\gamma|$, is controlled by the duration of the SDF pulse, $t$, and the phase of the displacement, $\mathrm{arg}(\gamma)$, is controlled by the SDF phase, $\phi_\mathrm{m}$. Following the SDF pulse, state measurement of the ancilla gives $\langle \hat{\sigma}_z \rangle =\mathrm{Re}[\langle \hat{D}(\gamma(t,\phi_\mathrm{m})) \rangle ]$, whilst the imaginary part can be obtained by applying an ancilla qubit rotation prior to the SDF. Setting $\gamma = \{\ell_\mathrm{s}, \ell_\mathrm{s} + i\ell_\mathrm{s}, i \ell_\mathrm{s}\}$ leads to measurements of $\{\langle \hat{X}_\mathrm{L}\rangle, \langle \hat{Y}_\mathrm{L}\rangle, \langle \hat{Z}_\mathrm{L}\rangle\}$. 

The expectation value of joint Pauli logical operators on two bosonic modes is obtained by applying two sequential SDF operations followed by an ancilla measurement, resulting in $\operatorname{Re}[\langle\hat{D}(\gamma) \otimes \hat{D}(\delta) \rangle ]$. 

To decode a GKP state, measuring Pauli expectation values of finite-energy GKP states with the operators of Eq.~\ref{eq:pauli_logical_op} is only valid in the limit $\Delta \rightarrow 0$. As explained in the main text, one can instead measure the expectation value of the Pauli measurement operators, which was done for the single-qubit gates and Bell state characterisation. These operators are defined as a summation over displacement operators on the logical GKP lattice~\cite{Shaw2024a}, 
\begin{subequations}
\begin{align}
     \hat{X}_\text{m}  &= \frac{1}{\pi} \sum_{n = -N}^{N-1} \frac{(-1)^n}{n+\frac{1}{2}}  \hat{D}\left(\ell_{\mathrm{s}}\left(n+1/2\right)\right), \\
     \hat{Y}_\text{m}  &= \frac{1}{\pi^2} \sum_{m,n = -N}^{N-1} \frac{ \hat{D}\left(\ell_{\mathrm{s}}(n+\frac{1}{2}) + i\ell_{\mathrm{s}}(m+\frac{1}{2})\right)}{(n+\frac{1}{2})(m+\frac{1}{2})}, \\
     \hat{Z}_\text{m}  &= \frac{1}{\pi} \sum_{m = -N}^{N-1} \frac{(-1)^m}{m+\frac{1}{2}} \;  \hat{D} \left(i\ell_{\mathrm{s}}\left(m+1/2\right)\right),
\end{align}
\label{eq:pauli_meas_op}
\end{subequations}
where $N$ sets a truncation, and the identity Pauli measurement operator $\hat{I}_\mathrm{m}$ coincides with the identity operator on Fock space. The expectation values $\langle \hat{X}_\mathrm{m} \rangle$, $\langle \hat{Y}_\mathrm{m} \rangle$ and $\langle \hat{Z}_\mathrm{m} \rangle$ are obtained by summing over expectation values of the displacement operator, $\langle \hat{D}(\gamma) \rangle$. Here, $\gamma$ is varied to sample the various points determined by Eq.~\ref{eq:pauli_meas_op}. Expectations of joint Pauli measurement operators for two bosonic modes, such as $\langle \hat{X}_\mathrm{m}\otimes\hat{Z}_\mathrm{m} \rangle $, can be obtained by taking the tensor product of any two single-mode Pauli measurement operator of Eq.~\ref{eq:pauli_meas_op}. Their joint expectation value will also be a summation over the expectation values of two-mode displacement operators.

The number of required measurements is reduced by using the fact that $\langle \hat{D}(\gamma)\rangle = \langle  \hat{D}(-\gamma)\rangle^*$. With this, the total number of measurements to calculate the three single-mode expectation values of the Pauli measurement operators of Eq.~\ref{eq:pauli_meas_op} is $N_\mathrm{meas} = 2 N + 2 N^2$. For the expectation values of the 15 non-trivial two-mode Pauli measurement operators, the number of measurements is $N_\mathrm{meas} = 4 N + 8 N^2 + 8 N^3 + 4 N^4$. The finite-energy envelope of the GKP states implies that $\langle \hat{D}(\gamma) \rangle \approx 0$ for large enough $|\gamma|$, meaning that the sums in Eq.~\ref{eq:pauli_meas_op} converge with $N$. In both the single-qubit gates and Bell state experiments, we choose a truncation of $N=2$ to minimise the number of measurements required, whilst still achieving a good approximation (in section~\ref{sm:imperfect_logical_meas} we quantify the error arising from this truncation). This results in $N_\mathrm{meas}=12$ and $N_\mathrm{meas}=168$ measurements for logical single-qubit and two-qubit states, respectively. For the ${\mathrm{CZ}}_\mathrm{L}$ gate experiment, we choose not to measure the Pauli measurement operators because the required number of measurements for all 16 states is $2688$, leading to a much longer experimental runtime. Instead, we choose to measure the expectation values of the logical Pauli operators of Eq.~\ref{eq:pauli_logical_op}, which requires a total of $240$ measurements, and incorporate the error from using these operators into our error budget (see Section~\ref{sm:imperfect_logical_meas}).

\subsection{Quantum Process Tomography}
\label{sm:qpt}

We use quantum process tomography (QPT) to characterise the action of the logical GKP gates. In general, QPT aims to reconstruct an unknown process from measurements on sets of input states and corresponding output states, the results of which can then be used to calculate a process fidelity. The input and output density matrices $\hat{\rho}_\mathrm{L}^\mathrm{in}$, $\hat{\rho}_\mathrm{L}^\mathrm{out}$ are reconstructed through a logical quantum state tomography (QST) routine, which consists of measuring expectation values of logical Pauli operators. The density matrices are related by a process $\mathcal{E}$, such that $\hat{\rho}_\mathrm{L}^\mathrm{out} = \mathcal{E} (\hat{\rho}_\mathrm{L}^\mathrm{in})$. 

In the single-qubit experiment, the logical density matrices $\hat{\rho}^\mathrm{in}_{\mathrm{L}}$ and $\hat{\rho}^\mathrm{out}_{\mathrm{L}}$ are reconstructed from Pauli measurement expectation values using the relation
\begin{equation}
    \label{eq:qpt_rho_1q}
    \hat{\rho}_\mathrm{L} =\frac{1}{2}\hat{E}^0 + \frac{1}{2} \sum_{i=1}^3 \langle \hat{E}^i_{\mathrm{m}} \rangle \hat{E}^{i},
\end{equation}
where $\hat{E}^i \in \{\hat{I}, \hat{\sigma}_x, \hat{\sigma}_y,\hat{\sigma}_z\}$ are the usual one-qubit Pauli operators and $\hat{E}^i_{\mathrm{m}} \in \{\hat{I}_\mathrm{m},  \hat{X}_\mathrm{m}, \hat{Y}_\mathrm{m},\hat{Z}_\mathrm{m}\}$ are the Pauli measurement operators, which can be obtained from the experiment using Eq.~\ref{eq:pauli_meas_op}. We construct a complete set of $\hat{\rho}^\text{in}_\mathrm{L}$ by preparing one of the four states $\{\ket{+Z_\mathrm{L}}, \ket{-Z_\mathrm{L}}, \ket{+X_\mathrm{L}}, \ket{+Y_\mathrm{L}}\}$ and taking measurements for each of the expectation values in Eq.~\ref{eq:qpt_rho_1q}, leading to a total of $4 \times 12 =48$ measurements (see section~\ref{sup:log_meas}). Reconstructing each $\hat{\rho}^\mathrm{out}_{\mathrm{L}}$ is performed in a similar manner, but with the gate applied to each initial state.

In the two-qubit experiment, the input and output density matrices are reconstructed from Pauli expectation values using the relation
\begin{equation}
    \label{eq:qpt_rho_2q}
    \hat{\rho}_\mathrm{L} =\frac{1}{4}\hat{E}^0 + \frac{1}{4} \sum_{i=1}^{15} \langle \hat{E}^i_{\mathrm{L}} \rangle \hat{E}^{i},
\end{equation} 
where $\hat{E}^i \in \{\hat{I},\hat{\sigma}_x, \hat{\sigma}_y,\hat{\sigma}_z\}^{\otimes 2}$ are the two-qubit Pauli operators and $\hat{E}^i_{\mathrm{L}} \in \{\hat{I}_\mathrm{L}, \hat{X}_\mathrm{L}, \hat{Y}_\mathrm{L},\hat{Z}_\mathrm{L}\}^{\otimes 2}$ are the logical Pauli operators of Eq.~\ref{eq:pauli_logical_op}. We first reconstruct each $\hat{\rho}^\mathrm{in}_\mathrm{L}$ by preparing one of the 16 possible input states from the set $\{ \ket{+Z_\mathrm{L}}, \ket{-Z_\mathrm{L}}, \ket{+X_\mathrm{L}}, \ket{+Y_\mathrm{L}}\}^{\otimes2}$ and measuring the 15 different non-trivial logical Pauli operators, resulting in $16\times 15 = 240$ measurements. Similarly, each $\hat{\rho}^\mathrm{out}_\mathrm{L}$ is retrieved by applying the $\mathrm{CZ}_\mathrm{L}$ gate after preparing one of the input states.

After experimentally reconstructing the sets of input and output density matrices, $\{\hat{\rho}_{\mathrm{L},i}^\mathrm{in}\}$ and $\{\hat{\rho}_{\mathrm{L},i}^\mathrm{out}\}$, we can characterise the process $\mathcal{E}$ that relates $\hat{\rho}_{\mathrm{L},i}^\mathrm{out} = \mathcal{E}(\hat{\rho}_{\mathrm{L}, i}^\mathrm{in})$ \cite{White2007}. To do this, we first express $\mathcal{E}$ in the operator-sum representation for each input state,
\begin{align}
    \label{eq:qpt_sum_op}
     \hat{\rho}^\mathrm{out}_{\mathrm{L},i} = \mathcal{E}(\hat{\rho}_{\mathrm{L},i}^\mathrm{in}) & = \sum_{m, n} \chi_{mn} \hat{E}^m \hat{\rho}_{\mathrm{L}, i}^\mathrm{in} \hat{E}^{n \dagger},
\end{align}
where $\hat{E}^m$ are the one- or two-qubit Pauli operators. In the following, we make use of the fact that an arbitrary state $\hat{\rho}$ can be expressed as a sum of basis states, $\hat{\rho}_j$, using the relation, $\hat{\rho} = \sum_j A_{i j} \hat{\rho}_j$, for some matrix of coefficients $A$. With this, we replace the left-hand side, $\hat{\rho}^\mathrm{out}_{\mathrm{L}, i}$, and right-hand side, $\hat{E}^m \hat{\rho}^\mathrm{in}_{\mathrm{L}, i} \hat{E}^{n\dagger}$, of Eq.~\ref{eq:qpt_sum_op} with sums of basis states,
\begin{align}
    & \hat{E}^m \hat{\rho}_{\mathrm{L},i}^\mathrm{in} \hat{E}^{n\dagger} = \sum_j \beta^{mn}_{ij}\hat{\rho}_j, \label{eq:qpt_sum_state_in} \\
     & \hat{\rho}_{L,i}^\mathrm{out}= \sum_j \lambda_{i j} \hat{\rho}_j, \label{eq:qpt_sum_state_out}
\end{align}
such that Eq.~\ref{eq:qpt_sum_op} becomes
\begin{equation}
    \sum_j \lambda_{i j} \hat{\rho}_j  = \sum_{j m n} \chi_{m n} \beta_{i j}^{m n} \hat{\rho}_j,
\end{equation}
where $\beta$ and $\vec{\lambda}$ are tensors of coefficients. This final expression allows us to relate $\vec{\lambda} = \beta \vec{\chi}$. The elements of $\beta$ are calculated from Eq.~\ref{eq:qpt_sum_state_in} using the experimentally reconstructed input density matrices $\hat{\rho}^\mathrm{in}_{\mathrm{L}, i}$ and choosing an appropriate basis set $\hat{\rho}_j$. Similarly, the elements of $\vec{\lambda}$ are calculated from Eq.~\ref{eq:qpt_sum_state_out} using the experimentally reconstructed output density matrices, $\hat{\rho}^\mathrm{out}_{\mathrm{L}, i}$. With this, $\chi$ can be retrieved through matrix inversion, $\vec{\chi} = \beta^{-1} \vec{\lambda}$. However, the resulting elements of $\chi$ can lead to an unphysical process $\mathcal{E}$. To ensure that the process is physical, we instead find the matrix $\chi$ that minimises $||\beta \vec{\chi} - \vec{\lambda}||_2$, where $||\cdot||_2$ is the $\ell_2$-norm~\cite{Bialczak2010,Flhmann2019}. We use a convex optimiser under the following constraints. We first set $\chi$ to be a Hermitian and non-negative definite matrix. We then ensure that $\mathcal{E}$ is trace preserving by constraining $\sum_{m,n} \chi_{m,n} \hat{E}_m \hat{E}_n^\dagger = \mathds{1}$.

\subsection{Logical Bell State Tomography}
\label{sec:qst}

The Bell state is characterised using logical QST, which aims to reconstruct the logical density matrix of the experimentally prepared state. In the Bell state experiment, we reconstruct the logical density matrix from 
\begin{equation}
    \label{eq:qst_rho}
    \hat{\rho}_\mathrm{Bell} =\frac{1}{4}\hat{E}^0 + \frac{1}{4} \sum_{i=1}^{15} \langle \hat{E}^i_{\mathrm{m}} \rangle \hat{E}^{i},
\end{equation}
where $\hat{E}^i \in \{\hat{I}, \hat{\sigma}_x, \hat{\sigma}_y,\hat{\sigma}_z\}^{\otimes 2}$ are the two-qubit Pauli operators and $\hat{E}^i_{\mathrm{m}} \in \{\hat{I}_\mathrm{m},  \hat{X}_\mathrm{m}, \hat{Y}_\mathrm{m},\hat{Z}_\mathrm{m}\}^{\otimes 2} $ are the Pauli measurement operators. The expectation values are calculated from Eq.~\ref{eq:pauli_meas_op}, where a truncation of $N=2$ results in 168 unique measurements. 

After obtaining $\hat{\rho}_\mathrm{Bell}$ from Eq.~\ref{eq:qst_rho}, we use a convex optimiser to obtain a physical density matrix, $\hat{\rho}$, which minimises $||\hat{\rho} - \hat{\rho}_\mathrm{Bell}||_2$ \cite{Riofro2017}. $\hat{\rho}$ is constrained to be non-negative definite and $\operatorname{Tr}(\hat{\rho}) = 1$. The resulting physical density matrix $\hat{\rho}$ is plotted in Fig.~\ref{fig:bell_data} of the main text.

\subsection{Error Analysis}

\begin{figure}
    \centering
   \includegraphics[]{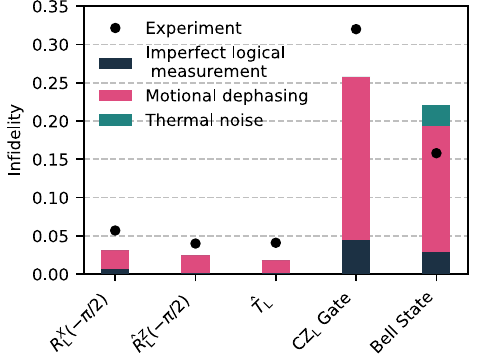}
    \caption{\textbf{Error analysis on the logical gates and Bell state preparation.} Simulated infidelities (bars) for various noise mechanisms are plotted alongside the experimentally measured infidelities (black circles). The included errors are those that are relevant to the actual experimental implementation. Imperfect logical measurement errors are obtained from the exact experimental routine used in the main text: single-qubit gates and the Bell state use SSSD with $N=2$, while the two-qubit gate measures the logical Pauli operators. We find in all cases that motional dephasing is the dominant error mechanism. Compared to the other process fidelity errors, the two-qubit gate error is particularly affected by imperfect logical measurement errors. We also observe that thermal noise is a significant error in the logical state fidelity of the Bell state.}
    \label{smfig:error}
\end{figure}

We analyse various error sources to identify the dominant noise mechanisms that limit the process and state fidelities reported in the main text. We consider the following errors: (i) imperfect logical measurement, (ii) motional dephasing, and (iii) thermal noise. In each case, fidelities are estimated by numerically simulating the exact experimental pulse sequences of Fig.~\ref{fig:circuits}. We apply the same numerical post-processing used for the experimental results to the simulation results for consistency. The results of which are summarised in Fig.~\ref{smfig:error}. We do not include errors from the heating of the bosonic modes as the heating rate of the ion trap is negligible (0.2 phonons/s). We find that motional dephasing during state preparation and application of the gate is the dominant error mechanism and accounts for most of the errors. Finally, we discuss a path toward reducing the motional dephasing error through realistic improvements to the experimental system. 

The error mechanisms, numerical simulations and resulting fidelities are discussed in detail below. 

\subsubsection{Imperfect logical measurements}
\label{sm:imperfect_logical_meas}

Imperfect logical measurements arise from the finite-energy approximation of the GKP states. First, this affects the reconstruction of the logical states, since expectation values of the logical Pauli operators on the approximate eigenstates, such as $\bra{+X_\mathrm{L}} \hat{X}_\mathrm{L} \ket{+X_\mathrm{L}} <1$, deviate from unit value. Second, fidelities estimated by QPT are also affected by the Gaussian envelope, which leads to unequal suppression of logical measurement readout values. This arises from the difference in displacement length between \(\langle \hat{Y}_\mathrm{L} \rangle\) measurements and those of \(\langle \hat{X}_\mathrm{L} \rangle\) or \(\langle \hat{Z}_\mathrm{L} \rangle\) \cite{Flhmann2019}. These errors can be mitigated by instead using the expectation value of the Pauli measurement operators through an SSSD routine at the cost of more experimental measurements (see section~\ref{sup:log_meas}). 

We estimate the infidelities from imperfect logical measurements by numerically simulating the QPT and QST processes on finite-energy GKP states. We obtain input and output GKP states from decoherence-free simulations of the experiment. Although this incorporates errors from the imperfections of the numerically optimised pulses, these errors are outlined in Section \ref{sup:num_op} and are found to not significantly affect the results detailed below. 

First, we simulate QPT for the single-qubit experiments by reconstructing the density matrix from Eq.~\ref{eq:qpt_rho_1q}. Commensurate with the experimental implementation, we measure expectation values of the Pauli measurement operators of Eq.~\ref{eq:pauli_meas_op} and set the truncation to $N=2$. The simulated logical process fidelities for the single-qubit gates $\{\hat{R}^X_\mathrm{L}(-\pi/2)$, $ \hat{R}^Z_\mathrm{L}(-\pi/2),\hat{T}_\mathrm{L}\} $ are $\{0.993, 0.998, 0.999\}$. 

Second, we simulate QPT for the two-qubit experiment. We reconstruct the density matrix from Eq.~\ref{eq:qpt_rho_2q} by computing expectations of the logical Pauli operators of Eq.~\ref{eq:pauli_logical_op}. We obtain a logical process fidelity of $0.955$. For consistency with the other experiments, we also report the process fidelity using expectation values of the Pauli measurement operators with a truncation of $N=2$, resulting in a process fidelity of 0.992.

Third, we simulate QST for the Bell state preparation experiment by reconstructing the density matrices from Eq.~\ref{eq:qst_rho}. We measure expectation values of the Pauli measurement operators with truncation $N=2$ and find a logical state fidelity of $0.971$. 

Finally, we investigate the errors due to the finite truncation of the sums for the Pauli measurement operators (Eq.~\ref{eq:pauli_meas_op}). We quantify the fidelities as we increase the SSSD truncation; we find that the fidelities converge at $N=3$ where we determine errors in the logical process fidelities of $\{2\times {10}^{-3}, 2\times {10}^{-5}, 4\times {10}^{-4} \}$ for the single-qubit gates and $5\times {10}^{-3}$ for the two-qubit gate. The error in the logical state fidelity for the Bell state is 0.012.

\subsubsection{Motional Dephasing}
\label{subsec:MotionalDepasing}

Motional dephasing arises from fluctuations of the motional mode frequencies, which result from noise in the ion trap's RF resonator and electronic circuit. These fluctuations are modelled by the noisy Hamiltonian
\begin{equation}
  \hat{H}^\mathrm{deph.}_j(t) = \epsilon_j(t) \hat{a}^\dagger_j \hat{a}_j,  
  \label{eq:H_dephasing}
\end{equation}
where $\epsilon_j(t)$ are fluctuations on mode $j$. 

\begin{figure*}
    \centering
   \includegraphics[]{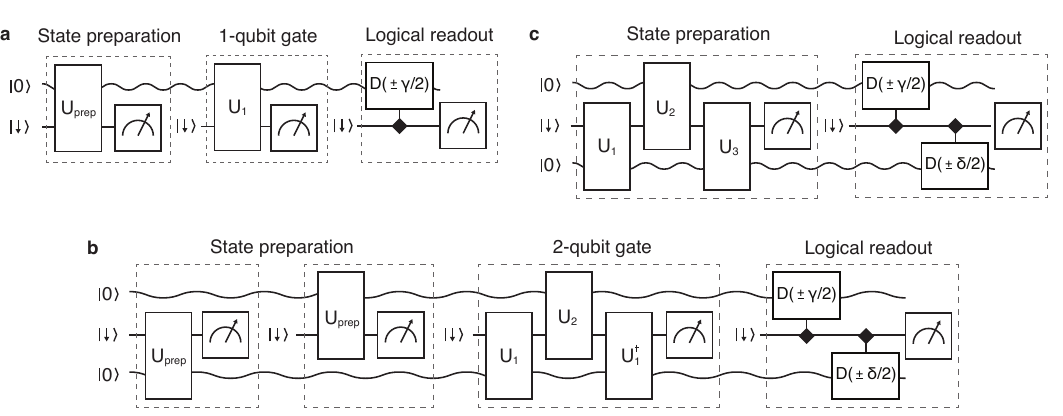}
    \caption{\textbf{Experimental pulse sequences.}
    \textbf{a}, single-qubit gate experiment with Quantum Process Tomography (QPT), \textbf{b}, two-qubit $\mathrm{CZ}_\mathrm{L}$ gate with QPT and \textbf{c}, Bell state preparation with Quantum State Tomography (QST). All unitary operators $\hat{U}_j$ are obtained from numerical optimisations, and are described by the state-dependent force Hamiltonian of Eq.~\ref{Eq:ControlHamiltonian}. $\hat{U}_\mathrm{prep}$ of \textbf{a} and \textbf{b} is chosen from one of four numerically optimised waveforms to prepare the GKP logical states $\ket{\pm Z_\mathrm{L}}$, $\ket{+X_\mathrm{L}}$ or $\ket{+Y_\mathrm{L}}$. $\hat{U}_1$ in \textbf{a} is one of three waveforms to perform the selected single-qubit gate. $\hat{U}_{1,2}$ and $\hat{U}_{1,2,3}$ of \textbf{b} and \textbf{c} are numerically optimised waveforms to implement a $\mathrm{CZ}_\mathrm{L}$ gate and prepare a Bell state, respectively. For the process tomography experiments of \textbf{a} and \textbf{b}, input states are reconstructed by witholding the respective gate operations and subsequent mid-circuit measurement. During logical readout, a single or two SDF pulses are applied to measure the single or joint-expectation value of the displacement operator. The SDF pulses apply a displacement conditioned on the qubit state in the $\hat{\sigma}_x$ basis (black diamond). Mid-circuit measurements are introduced after state-preparation and gate application to remove residual spin-motion entanglement. The experiment only proceeds if the $\ket{\downarrow}$ of the qubit state is measured, as photons scattered during an $\ket{\uparrow}$ measurement decohere the bosonic modes due to the recoil energy.     }
    \label{fig:circuits}
\end{figure*}

We estimate the impact of motional dephasing on the fidelity by including the dephasing terms of Eq.~\ref{eq:H_dephasing} in numerical simulations that incorporate the exact experimental pulse sequence as depicted in Fig.~\ref{fig:circuits}. For each experiment, we simulate QPT or QST by measuring the expectation values of Pauli operators as detailed in Sections~\ref{sm:qpt} and \ref{sec:qst}. We include dephasing during state preparation and gate application by replacing each unitary in Fig.~\ref{fig:circuits} with a noisy operation, $\hat{U}_j \rightarrow \hat{\tilde{U}}_j$, where $\hat{\tilde{U}}_j = \hat{\mathcal{T}}\mathrm{Exp}(- i \int^T_0 \hat{\tilde{H}}_j(t)dt)$. The SDF pulses during logical readout are modelled as error-free. $\hat{\tilde{H}}_j$ describes a noisy SDF pulse interacting with mode $j$ and is comprised of the modified SDF Hamiltonian of Eq.~\ref{eqn:ModifiedControlHamiltonian} and motional dephasing Hamiltonians of both modes, 
\begin{equation}
    \hat{\tilde{H}}_j(t) = \hat{H}_j(t) + \hat{H}^\mathrm{deph.}_1(t) + \hat{H}^\mathrm{deph.}_2(t).
    \label{eq:H_tot_noise}
\end{equation}
Pauli expectation values are measured under a random noise trajectory, $\epsilon_{1,2}(t)$. From these measurements, a density matrix is retrieved using Eqs.~\ref{eq:qpt_rho_1q}, \ref{eq:qpt_rho_2q} and \ref{eq:qst_rho} for the single-qubit, two-qubit and Bell state experiments, respectively. These simulations are repeated over $10^{3}$ noise realisations, from which we obtain an ensemble-averaged density matrix~\cite{Chenu2017, Gu2019}. The process fidelities for QPT and the state fidelity from QST are finally calculated from these averaged density matrices.

The noisy fluctuations $\epsilon_{1,2}(t)$ are modelled as piecewise constant functions. To simulate a noisy operator $\hat{\tilde{U}}_j$, the segment durations of $\epsilon_{1,2}(t)$ are set to be equal to the segment duration of the control pulse, $T_\mathrm{seg}$. The segment values of $\epsilon_{1,2}(t)$ are independent and identically distributed and are sampled from a normal distribution $\mathcal{N}(0, \sigma^2)$. The variance is calculated as $\sigma^2 = 2 \gamma / \tau$, where $\gamma = \SI{18}{Hz}$ is the decay rate~\cite{Matsos2024} and $\tau = T_\mathrm{seg}$ is the correlation time.

The simulated process fidelities under motional dephasing in the single-qubit gate experiments for the gates $\{\hat{R}^X_\mathrm{L}(-\pi/2)$, $ \hat{R}^Z_\mathrm{L}(-\pi/2),\hat{T}_\mathrm{L}\} $ are $\{ 0.969, 0.975, 0.982\} $. The simulated process fidelity for the CZ gate experiment is $0.742$, and the fidelity for the Bell state is $0.807$. We note that for the two-qubit gate, the error determined through the simulation also includes a significant error due to imperfect logical measurements. This measurement error can be reduced by reconstructing the states with the Pauli measurement operators. By performing the same simulation with an SSSD truncation at $N=2$, we determine a process fidelity of 0.795. 
This suggests a $\sim0.05$ improvement can be made to the $\mathrm{CZ}_\mathrm{L}$ result through an SSSD routine. Overall, these fidelities agree well with the experimentally measured fidelities, and we conclude that motional dephasing during preparation and gate application is the dominant source of error.

We additionally simulate the effects of motional dephasing during logical readout. We simulate the SDF pulses during logical readout as noisy operations evolving under the Hamiltonian of Eq.~\ref{eq:H_tot_noise}, and model all other unitaries as ideal. We find that the fidelities do not significantly differ from the fidelities of Section~\ref{sm:imperfect_logical_meas}  that only consider imperfect logical readout. 

\subsubsection{Thermal noise}

Thermal noise arises from imperfect cooling of the bosonic modes to the vacuum state. In our experiment, imperfect cooling arises from photon recoils, miscalibrations, and motional dephasing during sideband cooling. Infidelities from thermal noise are obtained by repeating the numerical simulations of section~\ref{sm:imperfect_logical_meas} and initialising the bosonic modes in a thermal state with an average phonon occupation of $\bar{n} = 0.05$. 

We find that fidelities in the single-qubit and two-qubit gate experiments do not significantly differ when adding thermal noise. However, we notice that an error of 0.028 is introduced in the prepared Bell state. 

\subsubsection{Improvements}

The logical process and state fidelities can be significantly improved by increasing the light-atom interaction strength ($\Omega_j$ of Eq.~\ref{eqn:ModifiedControlHamiltonian}) or by decreasing the motional dephasing rate. We repeat the numerical simulations under motional dephasing outlined in Section~\ref{subsec:MotionalDepasing} with an increased Rabi frequency of $\Omega_j = 2\pi \times  \SI{24}{kHz}$ and a reduced motional dephasing rate of $\gamma = \SI{5}{Hz}$. For each of the three experiments, we use the Pauli measurement operators through SSSD to reconstruct logical density matrices. We use an SSSD truncation of $N=2$ across all simulations. For the single-qubit gates, $\{\hat{R}^X_\mathrm{L}(-\pi/2)$, $ \hat{R}^Z_\mathrm{L}(-\pi/2),\hat{T}_\mathrm{L}\} $, we obtain logical process fidelities of \{0.994, 0.998, 0.999\}. For the $\mathrm{CZ}_\mathrm{L}$, we obtain a logical process fidelity of 0.987. The logical bell state fidelity is 0.968. In all three cases, the determined errors include contributions from imperfect logical measurements in the SSSD routine, see Section~\ref{sm:imperfect_logical_meas}. By comparing the fidelities with the $N=2$ truncation reported in Section~\ref{sm:imperfect_logical_meas}, we estimate that motional dephasing under improved experimental parameters contributes an additional error of up to $5\times 10^{-3}$ across all experiments.

\end{document}